\documentclass[a4paper,fleqn,usenatbib]{mnras}
\usepackage{newtxtext,newtxmath}
\usepackage[T1]{fontenc}
\usepackage{ae,aecompl}
\usepackage{graphicx}	
\usepackage{amsmath}	
\usepackage{amssymb}	
\usepackage{mathrsfs}

\title[Type~Ibn SNe~2006jc and 2015G]{Origins of Type~Ibn SNe~2006jc/2015G in interacting binaries and implications for pre-SN eruptions}

\author[N.-C. Sun et al.]{Ning-Chen Sun,$^{1}$\thanks{E-mail: n.sun@sheffield.ac.uk}
Jusytn R. Maund,$^{1}$
Ryosuke Hirai, $^{2}$
Paul A. Crowther, $^{1}$
\newauthor
and Philipp Podsiadlowski$^{2}$ \\
$^{1}$Department of Physics and Astronomy, University of Sheffield, Hicks Building, Hounsfield Road, Sheffield S3 7RH, UK\\
$^{2}$Department of Physics, University of Oxford, Keble Road, Oxford, OX1 3RH, UK}

\date{Accepted XXX. Received YYY; in original form ZZZ}

\pubyear{2019}

\begin{document}
\label{firstpage}
\pagerange{\pageref{firstpage}--\pageref{lastpage}}
\maketitle

\begin{abstract}
Type~Ibn supernovae (SNe~Ibn) are intriguing stellar explosions whose spectra exhibit narrow helium lines with little hydrogen. They trace the presence of circumstellar material (CSM) formed via pre-SN eruptions of their stripped-envelope progenitors. Early work has generally assumed that SNe~Ibn come from massive Wolf-Rayet (WR) stars via single star evolution. In this paper, we report ultraviolet (UV) and optical observations of two nearby Type~Ibn SNe~2006jc and 2015G conducted with the {\it Hubble Space Telescope} ({\it HST}) at late times. A point source is detected at the position of SN~2006jc, and we confirm the conclusion of Maund et al. that it is the progenitor's binary companion. Its position on the Hertzsprung-Russell (HR) diagram corresponds to a star that has evolved off the main sequence (MS); further analysis implies a low initial mass for the companion star ($M_2$~$\le$~12.3$^{+2.3}_{-1.5}$~$M_\odot$) and a secondary-to-primary initial mass ratio very close to unity ($q$~=~$M_2/M_1$~$\sim$~1); the SN progenitor's hydrogen envelope had been stripped through binary interaction. We do not detect the binary companion of SN~2015G. For both SNe, the surrounding stellar populations have relatively old ages and argue against any massive WR stars as their progenitors. These results suggest that SNe~Ibn may have lower-mass origins in interacting binaries. As a result, they also provide evidence that the giant eruptions commonly seen in massive luminous blue variables (LBVs) can also occur in much lower-mass, stripped-envelope stars just before core collapse.
\end{abstract}


\defcitealias{Maund2016}{M16}
\defcitealias{Shivvers2017}{S17}
\defcitealias{Foley2007}{F07}
\defcitealias{Pastorello2007}{P07}
\defcitealias{Pastorello2008}{P08}

\section{Introduction}
\label{intro.sec}

Type~Ibn supernovae (SNe~Ibn) are intriguing stellar explosions whose spectra exhibit narrow helium lines with no or very weak hydrogen lines (\citealt{Foley2007}, hereafter \citetalias{Foley2007}; \citealt{Pastorello2007, Pastorello2008}, hereafter \citetalias{Pastorello2007} and \citetalias{Pastorello2008}, respectively; \citealt{Hosseinzadeh2017}). Their progenitors have lost their hydrogen (and maybe also helium) envelopes and are embedded in dense, helium-rich circumstellar material (CSM). The narrow spectral lines arise when the fast SN ejecta interacts with the slow-moving CSM. Thus, SNe~Ibn provide a unique connection between the ``stripped-envelope" and the ``interacting" SN populations.

In general, there are two main channels to produce stripped-envelope SNe. \citet{Gaskell1986}, for example, proposed that their progenitors could be single, classical Wolf-Rayet (WR) stars, which are initially massive [$M_{\rm ZAMS}$~$\ge$~25~$M_\odot$ for solar metallicity or $M_{\rm ZAMS}$~$\ge$~30~$M_\odot$ for half-solar metallicity at zero-age main sequence (ZAMS)] and lose their envelopes via strong wind mass loss (\citealt{Crowther2007}; we shall hereafter refer to them as WR stars for simplicity. Note that some other objects also exhibit WR-like spectra, such as the very massive MS stars with $M_{\rm ZAMS}$~$\sim$~100~$M_\odot$ and the central stars of planetary nebulae; such objects are not considered in this work unless specified. We also do not consider those stars stripped in interacting binaries as WR stars, but only those that lose their envelopes via single-star evolution). Alternatively, the progenitors of stripped-envelope SNe could be lower-mass stars whose envelopes are removed through binary interaction \citep[e.g. Roche-lobe overflow or common-envelope evolution;][]{Podsiadlowski1992}. For SNe~Ibn, early work generally assumed that they come from massive WR stars via single star evolution \citepalias{Foley2007, Pastorello2007, Pastorello2008}; whether interacting binaries could be a viable progenitor channel remains an important but poorly investigated problem. The recent discovery of PS-12sk even suggests that this SN~Ibn may not arise from stellar core collapse, as it is located in the outskirts of an elliptical galaxy \citep{Sanders2013, Hosseinzadeh2019}.

Research of this problem is of particular importance for understanding the latest evolutionary stage of the stripped-envelope progenitors of SNe~Ibn. The CSM of SNe~Ibn is very dense (otherwise, the observed narrow helium lines cannot be produced) and cannot be formed by relatively steady wind mass loss. For example, the Type~Ibn SN~2006jc requires an extreme mass-loss rate of $\dot{M}$~$\sim$~10$^{-1}$~$M_\odot$~yr$^{-1}$, which is too high to be reconciled with stellar wind; instead, the CSM must be formed via the giant eruptions of the SN progenitors, which occur just before core collapse and are associated with extreme mass loss \citep{Smith2017}. This indicates that the progenitors of SNe~Ibn must become wildly unstable as they approach the end of their lives -- something that has not been included in standard stellar evolutionary models. Note that giant eruptions usually refer to the dramatic mass-loss events seen in LBVs, which are massive ($M_{\rm ZAMS}$~$\ge$~25~$M_\odot$), hydrogen-rich stars with significant instabilities \citep[e.g. see the review of][]{Smith2014c}; but in this paper we extend the concept of ``giant eruptions" (sometimes ``eruptions" for short) to describe the phenomena that have extreme mass-loss rates ($\dot{M}$~$\gtrsim$~10$^{-2}$~$M_\odot$~yr$^{-1}$), last from months to years or even shorter, and are often associated with optical brightening by several magnitudes. They are observationally similar to LBV giant eruptions, but they do not necessarily come from LBVs nor arise from the same physical mechanism(s). Also we distinguish the giant eruptions from other types of pre-SN mass loss that are less extreme (e.g. RSG superwinds, which have lower mass-loss rates and can blow for up to several thousand years; \citealt{Heger1997}); in this paper, we focus only on the most dramatic giant eruptions since the mass-loss events that give rise to the dense CSM of SNe~Ibn and, in particular, the detected pre-SN outburst of SN~2006jc (see below) are so similar to the LBV giant eruptions.

If SNe~Ibn are produced via the single star channel, they will provide a unique opportunity to understand the giant eruptions of WR stars. In the \citet{Conti1976} Scenario, WR stars are the descendants of LBVs after losing their hydrogen (and maybe helium) envelopes, and it is quite unexpected to see WR stars experience giant eruptions like LBVs. Observationally, eruptions from WR stars are also very rarely detected (e.g. HD~5980 in the Small Magellanic Cloud has been witnessed to undergo an LBV-like eruption in 1993--1994 and exhibit a WR-type spectrum during the quiescent phase; also this star is hydrogen-rich; \citealt{Barba1995}, \citealt{Koenigsberger2004}, \citealt{Hillier2019} and references therein). The research of SNe~Ibn has led to the speculation that some WR stars may still contain residual LBV-like instabilities, leading to pre-SN eruptions with significant mass loss \citepalias{Foley2007, Pastorello2007, Pastorello2008}.

If SNe~Ibn come from interacting binaries, pre-SN giant eruptions may even occur in stars of much lower initial masses. The possibility of giant eruptions in stars less massive than LBVs (i.e. with $M_{\rm ZAMS}$~$<$~25~$M_\odot$) has been discussed in the context of SN~2008S-like events or Intermediate-Luminosity Red Transients \citep[ILRTs;][]{Prieto2008, Bond2009, Thompson2009}.  Such events have much lower explosion energies than normal core-collapse SNe, and direct progenitor detections imply that their progenitors are dust-enshrouded with very low initial masses \citep[e.g. $\sim$10~$M_\odot$ for SN~2008S;][]{Prieto2008}. It has been argued that such events are not real SN explosions but the giant eruptions of their progenitors \citep[e.g.][]{Smith2009b, Bond2009, Berger2009}. If so, LBV-like stellar eruptions could extend down to much lower-mass stars than previously thought (yet, it remains unclear whether SN~2008-like events are {\it pre-SN} eruptions or not). SN~2008S-like events have also been suggested to be genuine SN explosions of super-AGB (asymptotic giant branch) stars \citep{Thompson2009, Botticella2009}. Thus, more observations are still needed to confirm the nature of such events (e.g. to see if their progenitors have disappeared or if their progenitors produce any eruptions and/or SN explosions again in the future). Some SNe~IIn-P seem to have significant pre-SN eruptions, and their inefficient $^{56}$Ni yields are consistent with progenitors of only $M_{\rm ZAMS}$~=~8--10~$M_\odot$; however, the possibility of massive progenitors of $M_{\rm ZAMS}$~$>$~25--30~$M_\odot$ cannot yet be entirely excluded \citep[e.g.][]{Sollerman1998, Chugai2004b, Mauerhan2013b}. SNe~Ibn, on the other hand, provide an alternative opportunity to address this issue, if they have lower-mass progenitors from interacting binary systems. Combined with the hydrogen-rich SN~IIn population, they can trace pre-SN giant eruptions in stars over a wider mass spectrum and with the presence/absence of hydrogen envelopes. This will provide important observational constraints for the theoretical efforts in understanding pre-SN eruptions \citep[e.g.][]{Quataert2012, Shiode2014, Smith2014, Fuller2017, Fuller2018, Chevalier2012}.

SNe~Ibn are very rare as they account for only 2.5~percent of all core-collapse SNe \citepalias{Pastorello2008}. Currently, only 31 cases have been classified as this type, and only three (SNe~2002ao, 2006jc and 2015G) occurred in the local universe \citep[with distances of $D$~$<$~30~Mpc;][]{Hosseinzadeh2017}. Thus, the nearby SNe~Ibn serve as important targets to investigate the progenitor channels for SNe~Ibn. Among them, SN~2006jc (host galaxy: UGC~4904) is the class prototype and has received much attention from astronomers. It is also the first SN for which a precursor outburst has been detected: in 2004, an optical transient (UGC~4904-V1) appeared at the SN position and remained visible for a few days \citepalias{Pastorello2007}. This outburst is very similar to the giant eruptions of LBVs. The most common interpretation is that SN~2006jc's progenitor was a WR star which erupted two years before its terminal explosion \citepalias{Foley2007, Pastorello2007, Pastorello2008}. An alternative speculation invokes an LBV+WR binary system, in which the two stars produced the outburst and the SN respectively \citepalias{Pastorello2007, Pastorello2008}.

\citet[][hereafter \citetalias{Maund2016}]{Maund2016} detected a late-time source at the position of SN~2006jc. The SN itself had faded significantly by the time of their observations (April 2010), and the late-time source is most likely to be a hot binary companion of its progenitor. This makes it one of only four SNe with a binary companion detection (after SN~1993J, \citealt{Maund2004, Fox2014}; SN~2001ig, \citealt{Ryder2018}; and SN~2011dh \citealt{Folatelli2014, Maund2015}). It is also compelling evidence that SN~2006jc was produced in a binary system; however, no follow-up observations have been reported to confirm the nature of this late-time source (e.g. by seeing whether its brightness has changed since then). The 2010 optical photometry of the companion was only able to constrain the temperature to log($T_{\rm eff}$/K)~$>$~3.7. As a result, its properties were not tightly constrained and it remains unknown whether there has been any interaction between the companion star and the SN progenitor. Furthermore, SN~2006jc was noticed to occur in a sparse region of its host galaxy, located in the vicinity of a clump of young stars but with a clear offset. Its environment may contain important information of its progenitor system, but a more quantitative analysis has not been carried out.

SN~2015G (host galaxy: NGC~6951) is another nearby SN~Ibn and  \citet[][hereafter \citetalias{Shivvers2017}]{Shivvers2017} has made a comprehensive study on it. It shares many spectral features with SN~2006jc (e.g. narrow helium lines and a blue pseudo-continuum), and its CSM was formed in the last year or so before core collapse. Using pre-explosion images, \citetalias{Shivvers2017} found that its progenitor is not massive enough to be consistent with a WR star. Meanwhile, the surrounding stellar populations have relatively old ages and argue against any massive WR star as its progenitor. Thus, their analysis suggests that SN~2015G may have a lower-mass progenitor from an interacting binary system; however, a search for its binary companion has not been performed with late-time observations.

In this paper, we report new UV and optical observations of SNe~2006jc and 2015G conducted at late times with the {\it Wide Field Camera 3} ({\it WFC3}) of the {\it HST}. Combined with archival optical observations, we try to detect and infer the properties of their binary companions, and to analyse the surrounding stellar populations in their environments. Our aim is to understand whether SNe~Ibn come from massive WR stars through single-star evolution or from lower-mass progenitors in interacting binary systems.

Throughout this paper, we assume half-solar metallicity for both SNe; a distance of 27.8~Mpc is adopted for SN~2006jc and 23.2~Mpc for SN~2015G (all consistent with \citetalias{Maund2016} and \citetalias{Shivvers2017}). Both SNe were discovered when their brightness had already been on the decline. We assume that their peak brightness took place on 2006 Oct~1 for SN~2006jc and on 2015 Mar~4 for SN~2015G, according to the estimates by \citetalias{Pastorello2008} and \citetalias{Shivvers2017}, respectively. This paper is structured as follows. We describe the observations in Section~\ref{obs.sec}. In Section~\ref{source.sec} we report the detection or non-detection of any late-time sources at the SN positions, and in the case of a detection, discuss the possibility of the late-time source being a binary companion. We next try to constrain the physical properties of the companion star in Section~\ref{companion.sec}, and to investigate the SN's progenitor channel, when possible. We present an environment analysis of the two SNe in Section~\ref{env.sec}. In Section~\ref{discussion.sec} is a discussion on the progenitor channels and pre-SN eruptions of SNe~Ibn. We finally close this paper with a summary of our conclusions.

\section{Observations}
\label{obs.sec}

\begin{table*}
\label{log.tab}
\center
\caption{{\it HST} observations of SNe~2006jc and 2015G.}
\begin{tabular}{ccccccc}
\hline
\hline
Target & Date & Time from peak & Instrument & Filter & Exposure & Program\\
 & (UT) & brightness (yr) &   &    & Time (s) & ID \\
\hline
SN~2006jc & 2010 Apr 30.5 & 3.6 &  {\it ACS/WFC1} &  {\it F658N} & 1380 & 11675$^{\rm a}$ \\
 & 2010 Apr 30.6 & 3.6 &  {\it ACS/WFC1} &  {\it F625W} & \ 897 & 11675$^{\ }$ \\
 & 2010 Apr 30.6 & 3.6 &  {\it ACS/WFC1} &  {\it F555W} & \ 868 & 11675$^{\ }$ \\
 & 2010 Apr 30.6 & 3.6 &  {\it ACS/WFC1} &  {\it F435W} & \ 868 & 11675$^{\ }$ \\
 & 2017 Feb 25.8 & 10.4 &  {\it WFC3/UVIS2} &  {\it F300X} & 1200 & 14762$^{\rm a}$ \\
 & 2017 Feb 25.8 & 10.4 &  {\it WFC3/UVIS2} &  {\it F475X} & \ 350 & 14762$^{\ }$ \\
\hline
SN~2015G & 2015 Nov 25.6 & 0.7 & {\it WFC3/UVIS} & {\it F814W} & 780 & 14149$^{\rm b}$ \\
 & 2015 Nov 25.6 & 0.7 & {\it WFC3/UVIS} & {\it F555W} & 710 & 14149$^{\ }$ \\
 & 2016 Dec 25.4 & 1.8 & {\it WFC3/UVIS2} & {\it F300X} & 1200 & 14762$^{\ }$ \\
 & 2016 Dec 25.4 & 1.8 & {\it WFC3/UVIS2} & {\it F475X} & 350 & 14762$^{\ }$ \\
\hline
\multicolumn{7}{l}{a: PI: J. R. Maund} \\
\multicolumn{7}{l}{b: PI: A. Filippenko} \\
\end{tabular}
\end{table*}

We obtained new {\it HST} observations of SNe~2006jc and 2015G at late times, which are part of the Cycle~24 programme ``{\it A UV census of the sites of core-collapse SNe}'' (Program~14762; PI: J.~R.~Maund). The observations were acquired with the {\it ultraviolet-visible channel} (i.e. {\it UVIS}) of {\it WFC3}. The extremely wide UV filter, {\it F300X}, and the extremely wide blue filter, {\it F475X}, were used, and observations in each filter were composed of two separate dithered exposures. The images still suffer from a high level of cosmic-ray contamination after the standard calibration pipeline. Thus, we manually combined the exposures with the ASTRODRIZZLE package\footnote{\url{http://drizzlepac.stsci.edu/}}. In practice, the drizzle output pixel scale was set to 0.04~arcsec, matching the original pixel size. We found that using \texttt{driz\_cr\_grow~=~3} could remove the cosmic rays most efficiently, especially in the long-exposure {\it F300X} images. This setting adopts a larger radius than default around each detected cosmic ray within which to apply more stringent criteria for additional cosmic ray detection. Meanwhile, all other drizzle parameters were kept unchanged as in the standard {\it HST} calibration pipeline.

We also make use of archival {\it HST} observations. They include observations of SN~2006jc in April 2010 (Program~11675; PI: J.~R.~Maund) and of SN~2015G in November 2015 (Program~14149; PI: A.~Filippenko). The former set of observations were obtained with the {\it Advanced Camera for Surveys} ({\it ACS}), and their data reduction has been detailed in \citetalias{Maund2016}. The latter set of observations were acquired with {\it WFC3}; their images have been well processed by the standard calibration and are used without any further manual reduction. Table~\ref{log.tab} presents a summary of all the observations used in this work.

Photometry was conducted with the DOLPHOT package\footnote{\url{http://americano.dolphinsim.com/dolphot/}} \citep{Dolphin2000} with the {\it ACS} and {\it WFC3} specific modules. A full description of the adopted DOLPHOT parameters is presented in Appendix~\ref{dp.sec}. For SN~2006jc, a systematic error ($\sim$0.46~mag) was found in the raw {\it WFC3/F475X} magnitudes reported by DOLPHOT. We corrected this systematic error with a method described in Appendix~\ref{syserr.sec}.

\section{Late-Time Sources}
\label{source.sec}

\begin{figure*}
\centering
\includegraphics[scale=0.36, angle=0]{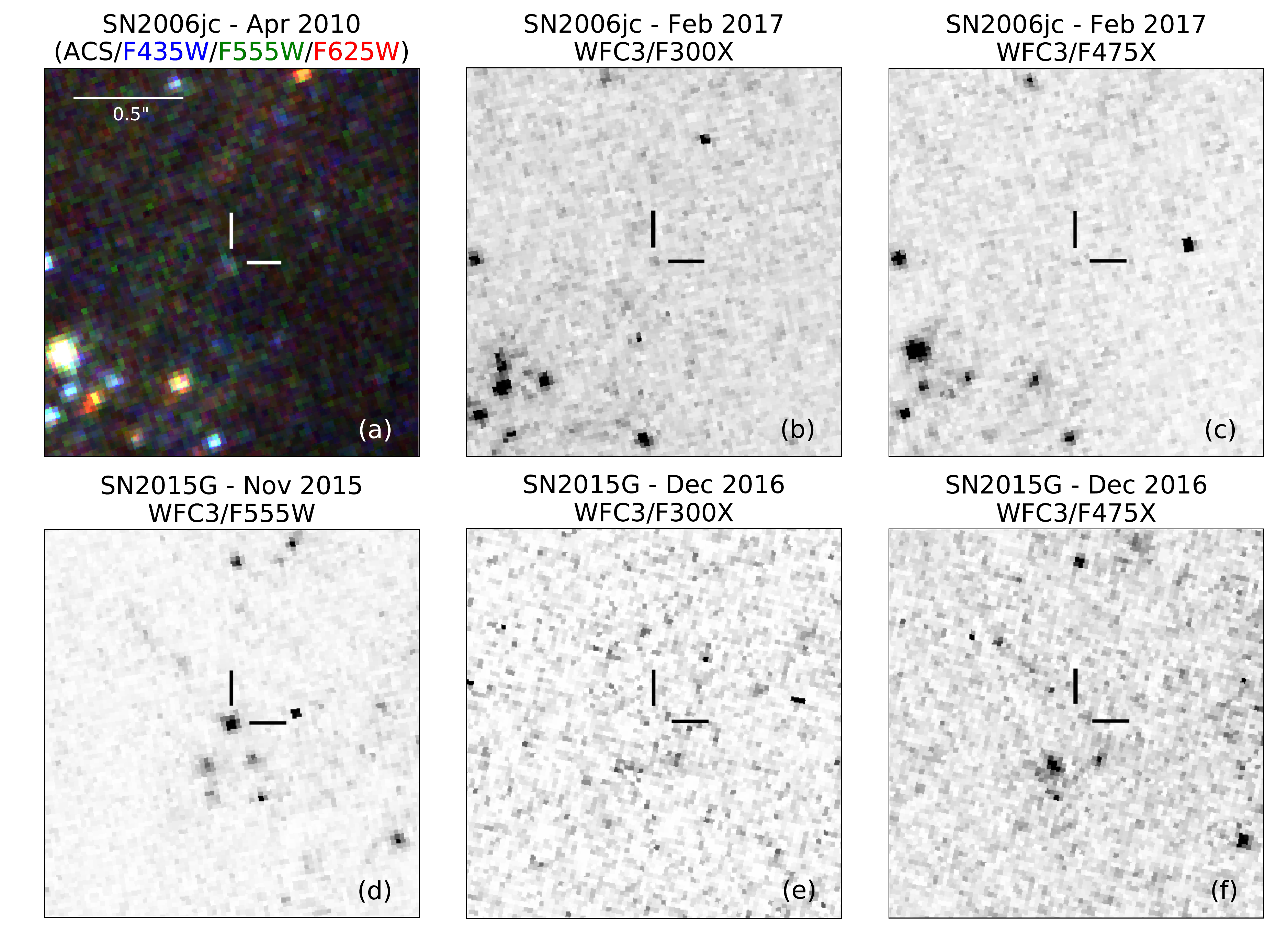}
\caption{{\it HST} observations of the sites of  SN~2006jc (top row) and SN~2015G (bottom row). The first panel is a three-colour composite of the {\it F435W}, {\it F555W}, and {\it F625W} images taken by {\it ACS} in 2010, while the other panels are single-band images from {\it WFC3} (as labelled). The crosshair in each panel corresponds to the SN position, and all the images have been set to the same angular scale and aligned with north up and east to the left.}
\label{image.fig}
\end{figure*}

\subsection{SN~2006jc}
\label{source06jc.sec}

Figure~\ref{image.fig} (top panels) shows the site of SN~2006jc as observed by {\it HST}. Using the {\it ACS} observations in 2010 (Fig.~\ref{image.fig}a), \citetalias{Maund2016} reported the detection of a late-time source at the SN position. In this work, we re-derived its magnitude (all in the Vega magnitude system throughout this paper) to be $m_{F435W}$~=~26.59~$\pm$~0.21~mag (${\rm S/N}$~=~5.1), $m_{F555W}$~=~26.51~$\pm$~0.21~mag (${\rm S/N}$~=~5.1), and $m_{F625W}$~=~26.55~$\pm$~0.23~mag (${\rm S/N}$~=~4.7). The magnitudes are within the photometric uncertainties from those reported by \citetalias{Maund2016}, and the slight differences arise from the different parameter settings in the photometry. On the other hand, this source is not detected in the narrow-band {\it ACS/F658N} observation to a significance of ${\rm S/N}$~$\ge$~3.0. The detection limit is estimated to be $m_{F658N}$~=~24.35~$\pm$~0.32~mag (also consistent with \citetalias{Maund2016}), based on artificial star tests at the SN position. An artificial star is considered to have been successfully recovered if it is detected at S/N~$\geq$~3 and within 1 pixel of the inserted position. The magnitude at which the detection probability falls to 50\% is regarded as the detection limit.

The {\it WFC3} observations in 2017 (i.e. $\sim$10 years after the SN explosion; Fig.~\ref{image.fig}b, c) also reveal a late-time source at the SN position with magnitudes of $m_{F300X}$~=~25.93~$\pm$~0.22~mag (${\rm S/N}$~=~5.0) and $m_{F475X}$~=~26.81~$\pm$~0.27~mag (${\rm S/N}$~=~4.1). Note that a systematic error has been corrected from the raw {\it WFC3/F475X} magnitude reported by DOLPHOT (see Appendix~\ref{syserr.sec} for details). We aligned the {\it ACS/F435W} and the {\it WFC3/F300X} images with 29 common stars, reaching a precision of 0.34 pixel (the misalignment between different filters of the same epoch are very small). The positions of the {\it WFC3} source and the {\it ACS} source differ by only 0.31 pixel and agree with each other within the astrometric uncertainties. Thus, both late-time sources are spatially coincident and most likely arise at the SN position.

\begin{figure*}
\centering
\includegraphics[scale=0.8, angle=0]{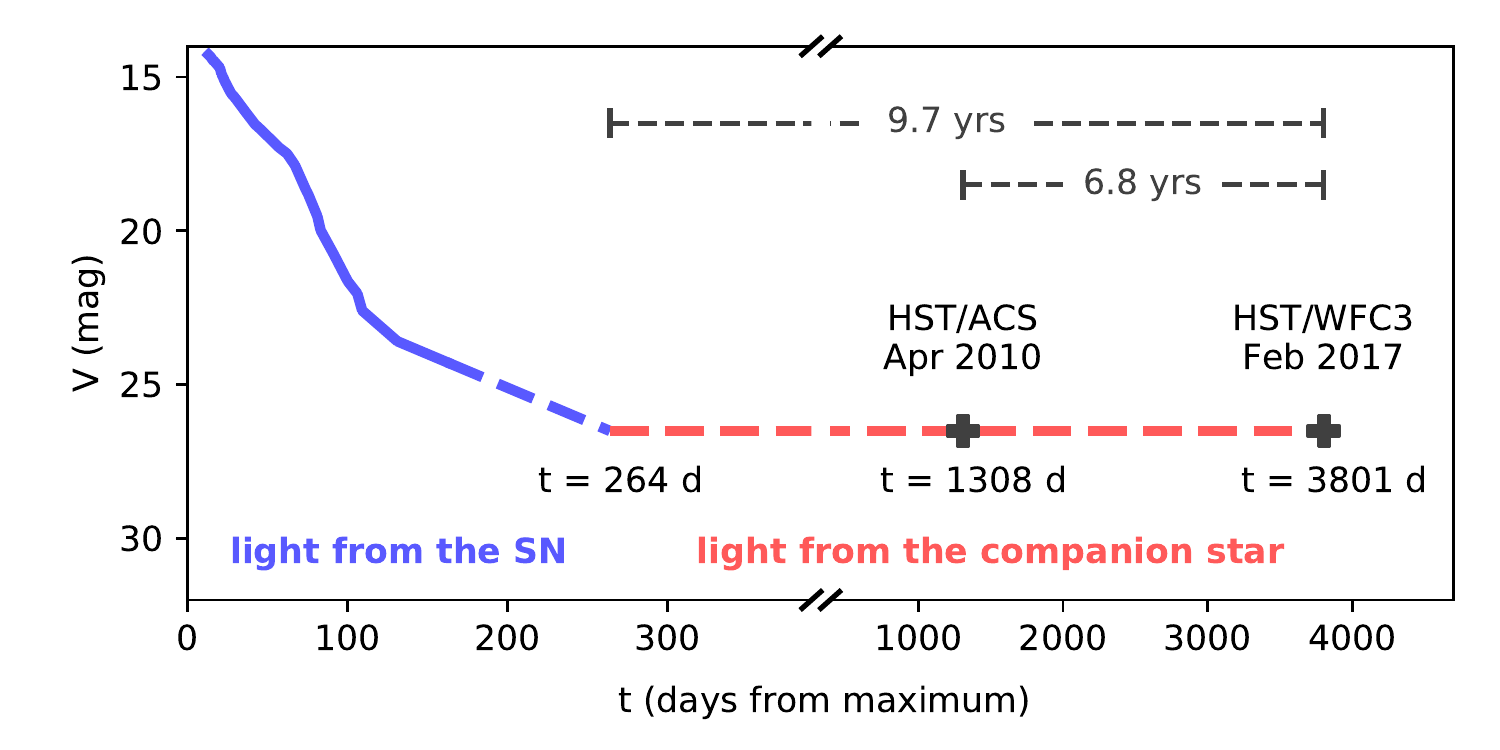}
\caption{$V$-band magnitude evolution of the source at the position of SN~2006jc. The solid line is the SN's light curve from \citetalias{Pastorello2007} and \citetalias{Pastorello2008}, and the two thick plus symbols are from the HST observations at late times. The dashed line is a possible evolution scenario if one assumes the light curve to have followed the trend as in \citetalias{Pastorello2007} and \citetalias{Pastorello2008} before it suddenly flattened to a constant level. The early light curve (blue) is dominated by the SN itself, and the long plateau at late times (red) is best explained by a binary companion.}
\label{curve06jc.fig}
\end{figure*}

\subsubsection{Nature of the late-time source}
\label{nature.sec}

The nature of the late-time source could have several possibilities. \citetalias{Maund2016} have argued that it is most likely to be a binary companion of SN~2006jc's progenitor star (see their Section~3 for a thorough discussion). With observations at a new epoch, we can reinforce this conclusion by excluding a light echo or new ejecta-CSM interaction at late times. Figure~\ref{curve06jc.fig} shows the $V$-band light curve of the source at the position of SN~2006jc. The solid curve is from \citetalias{Pastorello2007} and \citetalias{Pastorello2008} out to $t$~=~162~day after maximum, while the two thick plus symbols correspond to the {\it ACS} and {\it WFC3} observations in Apr 2010 and Feb 2017, respectively. For hot sources (which is true for this late-time source, see Section~\ref{companion.sec}), both the {\it ACS/F555W} and the {\it WFC3/F475X} magnitudes are roughly equal to the Johnson $V$-band magnitude \citep{Sirianni2005, Saha2011, Harris2018}. Considering the photometric uncertainties, the transformed $V$-band magnitudes are not significantly different between the two epochs. As a result, we have adopted a value of $V$~=~26.5~mag for both epochs in the figure.

It is immediately obvious that this late-time source has a very stable optical brightness over a long period (6.8~years) between the two epochs. The SN site may have remained at this brightness for an even longer period, if one assumes the light curve to have followed the trend as in \citetalias{Pastorello2007} and \citetalias{Pastorello2008} before it suddenly flattened and remained constant from then on. In this case, the flattening occurred at $t$~=~162~day after maximum, and the light curve should have stayed constant for at least 9.7~years. Such a long plateau is best explained by a companion star, whose brightness became dominant at late times after the SN itself had faded significantly.

Some SNe are surrounded by CSM at far distances which may interact with the ejecta at late times (e.g., SN~1988Z, \citealt{Turatto1993}; SN~1993J, \citealt{Zhang2004}; SN~2005ip, \citealt{Smith2009}, \citealt{Stritzinger2012}; SN~2009ip, \citealt{Mauerhan2013}; SN~2010mc, \citealt{Smith2014b}; SN~2014C, \citealt{Margutti2017}). Such new ejecta-CSM interaction provides an additional energy contribution, which may lead to a re-brightening, a slower decline rate or even a plateau in the light curve. However, plateaus produced in this way are very difficult to last for 6.8--9.7~years, since the interaction strength may change with varying CSM properties as the shock front propagates outwards. Only the extremely enduring SN~1988Z has ever been observed to have a comparably long plateau in its H$\alpha$ light curve \citep[which is a good proxy for interaction strength;][]{Smith2017b}. Yet, SN~1988Z belongs to the Type~IIn class and its progenitor may have had a very different mass-loss history from that of SN~2006jc. Thus, we suggest that SN~2006jc's late-time source is less likely to be caused by ejecta-CSM interaction arising at late times.

The observed late-time source is also inconsistent with a light echo reflected by either circumstellar or interstellar dust. \citet{Mattila2008} detected newly formed dust within 1000~AU, as well as a pre-existing dust shell extending to 1~pc from the SN. These two dust components are, however, too close to produce an observable light echo by the time of the {\it WFC3} observations in 2017. Meanwhile, SN~2006jc occurred in the outskirt of its host galaxy, where there is little interstellar dust \citepalias[][supplementary material]{Pastorello2007} which could scatter its light. In addition, it is very difficult to find a scattering dust configuration to produce a constant and unresolved light echo for SN~2006jc. For instance, in the case of a foreground dust sheet~\footnote{In the case of background dust, the scattering angles are larger than 90$^{\circ}$, which are inefficient \citep{Draine2003}.}, the light echo brightness remains constant only if $d$~$\gg$~$ct$, where $d$ is the distance between the dust sheet and the SN, $c$ is the speed of light, and $t$ is the time since the SN's radiation burst \citep{Chevalier1986, Cappellaro2001}. So if the observed late-time source is the light echo of SN~2006jc at 2010 and 2017, we derive a lower limit of $d$~$>$~240~light years in order to keep their magnitudes in agreement with each other within 0.3~mag. By the time of the 2017 observations, the corresponding ring-like light echo would have reached a radius of $\rho$~=~$\sqrt{ct(ct+2d)}$~$>$~0.16$\arcsec$. Such a light echo should be spatially resolvable by ${\it WFC3}$, whose point spread function has a full width at half maximum of only 0.07$\arcsec$ at optical wavelengths; however, the observed late-time source is point-like and contradicts this estimate. Thus, the late-time source is not likely to be produced by a light echo of SN~2006jc.

This late-time source, as \citetalias{Maund2016} pointed out,  is too faint to be consistent with an unresolved star cluster. It is also unlikely to be an unrelated star in chance alignment with SN~2006jc, since the location is in a very sparse area of its host galaxy (\citetalias{Maund2016} estimated the probability of chance alignment to be only $\sim$1\%). These conclusions remain unaltered with the new observations at 2017. Thus, this late-time source is most likely to be the SN progenitor's binary companion.

\subsection{SN~2015G}
\label{source15g.sec}

Figure~\ref{image.fig}d shows a {\it WFC3/F555W} image of SN~2015G in Nov 2015. The SN was still very bright at this epoch, and we were able to determine its position on this image with DOLPHOT. The SN position was then transformed onto the late-time {\it WFC3/F475X} image using 20 common stars. No late-time source was detected significantly (S/N~$\geq$~3) at the SN site in either the {\it WFC3/F300X} (Fig.~\ref{image.fig}e) or the {\it WFC3/F475X}  (Fig.~\ref{image.fig}f) band. The nearest source is 2.7~pixels away from the transformed SN position, much larger than the uncertainty (0.23~pixel). This source is point-like and very faint (only marginally detected in the {\it WFC3/F475X} band with $m_{F475X}$~=~27.43~$\pm$~0.30~mag). Thus, it is unlikely to be a star cluster but rather an unrelated star near SN~2015G. With artificial star tests, we determined the detection limits (to S/N~$\geq$~3) of the late-time observations to be $m_{F300X}$~=~26.01~$\pm$~0.17~mag and $m_{F475X}$~=~27.74~$\pm$~0.19~mag.

\section{Binary Companions}
\label{companion.sec}

\subsection{SN2006jc}
\label{companion06jc.sec}

\begin{figure*}
\centering
\includegraphics[scale=0.46, angle=0]{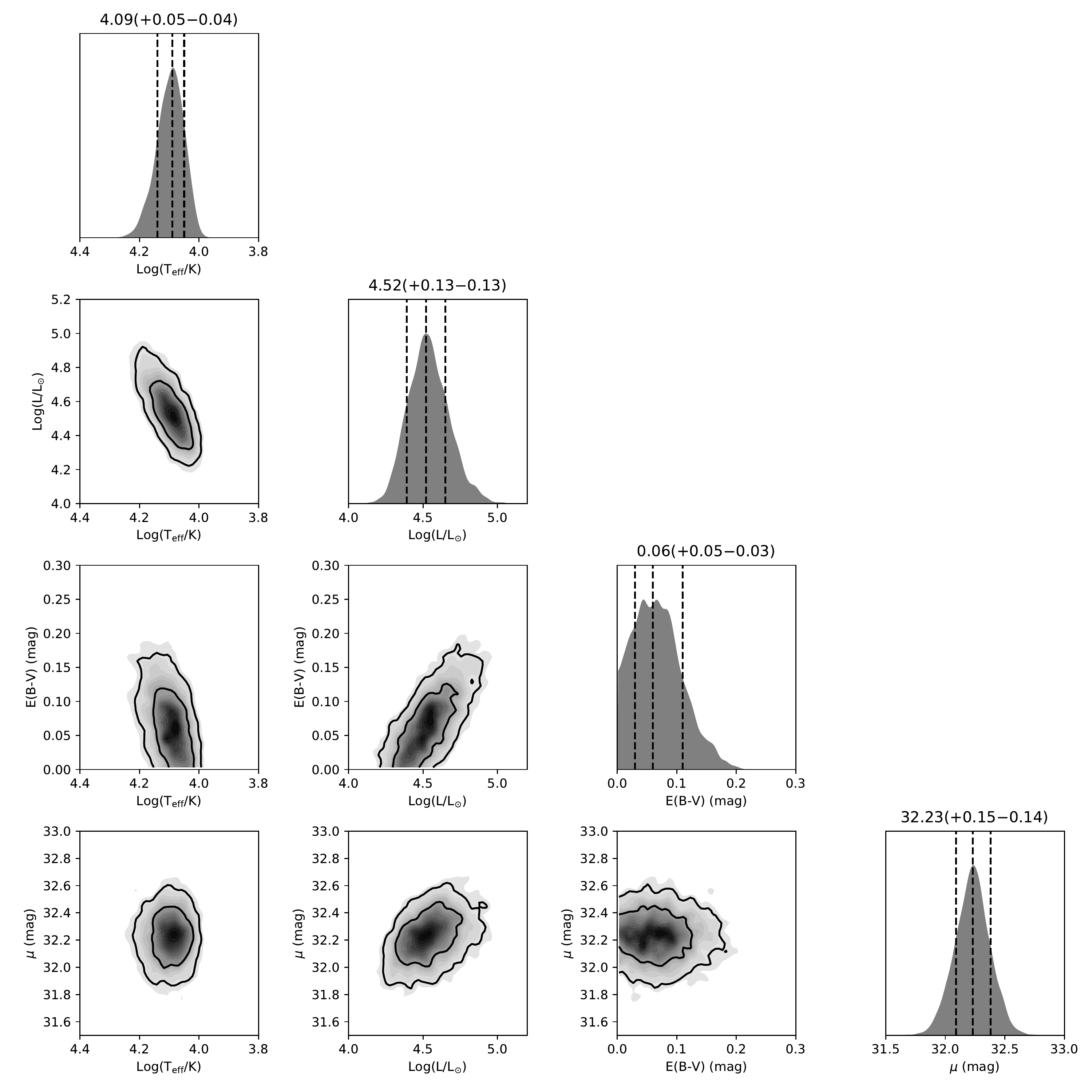}
\caption{Posterior probability distributions of effective temperature, luminosity, interstellar reddening, and distance modulus. The numbers on top of and the dashed lines in each histogram show the median value with the 68\% (1-sigma) credible interval of the marginalised posterior probability. The contours in the other panels, from inside to outside, contain 68\% and 95\% marginalised posterior probability.}
\label{corner06jc.fig}
\end{figure*}

\begin{figure*}
\centering
\includegraphics[scale=0.75, angle=0]{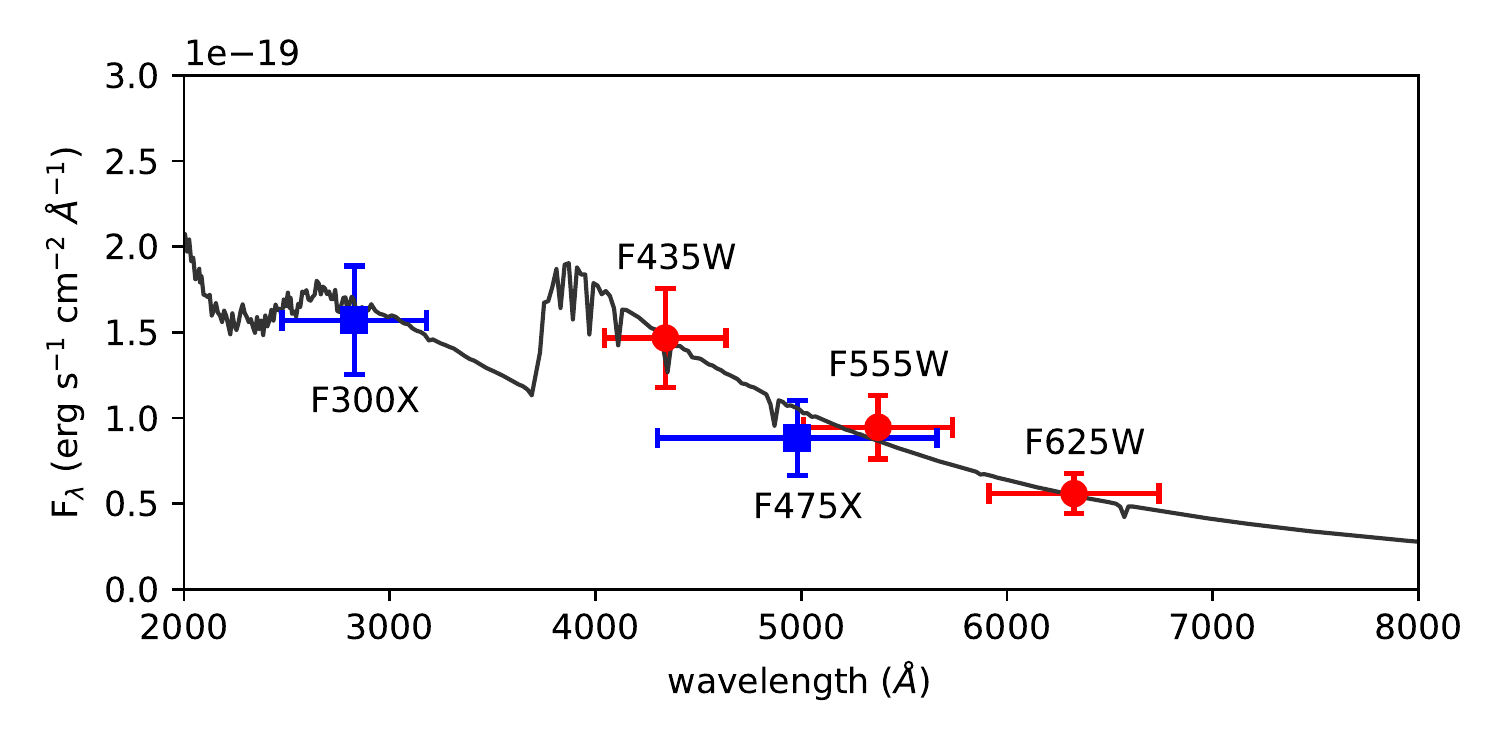}
\caption{The observed SED of the late-time source at the position of SN~2006jc, as acquired by {\it ACS} in April 2010 (red circles) and by {\it WFC3} in February 2017 (blue squares); the vertical error bars show the photometric errors, while the horizontal error bars represent the root-mean-square band widths of the filters. Overlaid is the best-fitting supergiant spectrum. }
\label{spec06jc.fig}
\end{figure*}

\begin{figure}
\centering
\includegraphics[scale=0.8, angle=0]{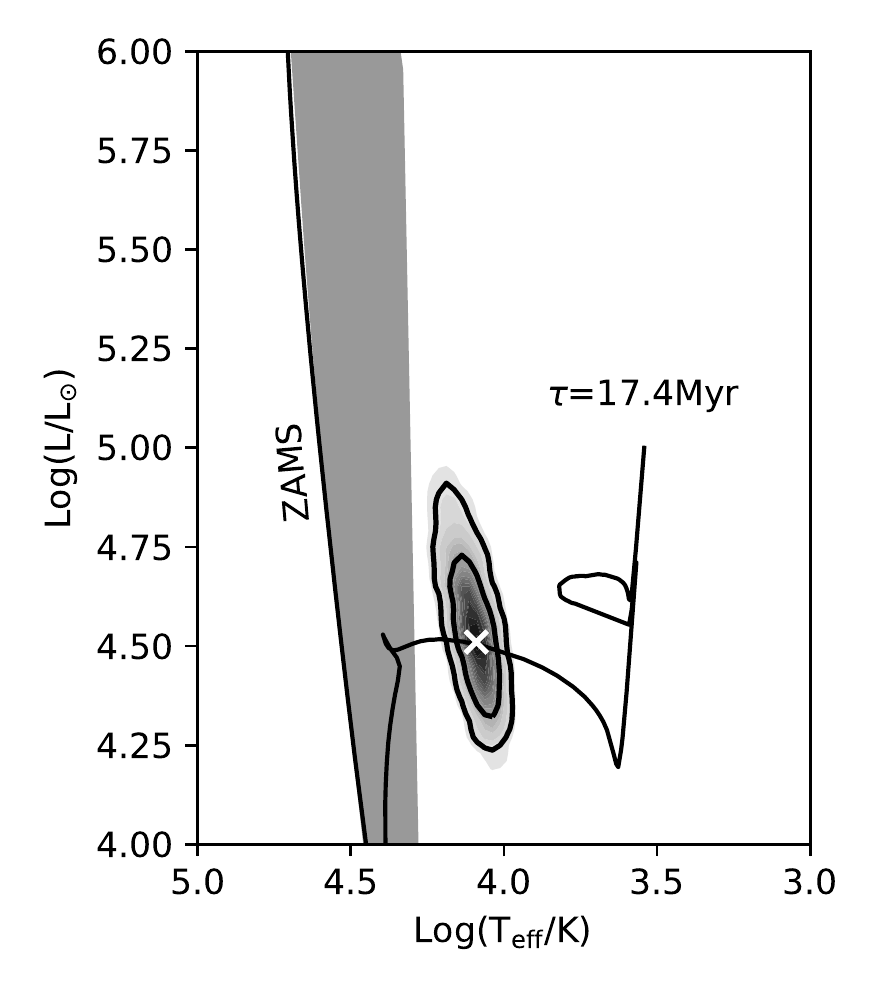}
\caption{Position of SN~2006jc's binary companion on the HR diagram. The `$\times$' symbol indicates its most likely solution, while the two contours from inside to outside contain 68\% and 95\% of the probability, respectively. Overlaid is a PARSEC (v1.2S) stellar isochrone of 17.4~Myr and half-solar metallicity. The grey shaded region along the ZAMS indicates the locus of binary companions at the death of the primary as predicted by the BPASS models.}
\label{hrd06jc.fig}
\end{figure}

We try to fit the observed spectral energy distribution (SED) with ATLAS9 synthetic spectra \citep{Castelli2004} for single supergiants\footnote{The influence of surface gravity on broad-band photometry is very small compared with the photometric uncertainties (especially for hot sources).} in order to derive the physical properties of SN~2006jc's companion. We assume a Gaussian prior for distance modulus, $\mu$, based on the reported distance value of 27.8~$\pm$~1.9~Mpc \citepalias[i.e. $\mu$~=~32.22~$\pm$~0.15~mag;][]{Maund2016}. \citetalias{Pastorello2007} estimated a total interstellar reddening of $E(B-V)$~=~0.05~mag for SN~2006jc; we adopt this value and conservatively assumed that it has an error of 0.05~mag. The prior in reddening is also assumed to be Gaussian, but truncated at zero to avoid negative reddening values. A Galactic extinction law with $R_V$~=~3.1 \citep{Cardelli1989} is used in the analysis\footnote{It is reasonable to assume this extinction law given the low interstellar reddening. Changing the extinction law does not affect the results. For example, $E(B-V)$~=~0.05~mag corresponds to $A_{F300X}$~=~0.31~mag with the assumed extinction law (for the most affected UV band); if we change $R_V$ to 5.0 or 2.1, $A_{F300X}$ becomes 0.36 and 0.28~mag, respectively, which are very similar to the original value. The changes in extinction are no larger than 0.05~mag and are much smaller than the photometric uncertainties. We also repeated the SED fitting with different extinction laws, and the results are very similar to those given in this section.}. Figure~\ref{corner06jc.fig} shows a corner plot of the posterior probability distributions. It can be seen that the uncertainties of $\mu$ and $E(B-V)$ have larger effect on the luminosity but smaller influence on the effective temperature.

The SED fitting suggests that the companion star has an effective temperature of log($T_{\rm eff}$/K)~=~4.09$^{+0.05}_{-0.04}$ and a luminosity of log($L/L_\odot$)~=~4.52$^{+0.13}_{-0.13}$ (the best-fitting spectrum is shown in Fig.~\ref{spec06jc.fig}). Figure~\ref{hrd06jc.fig} displays the star's possible position on the Hertzsprung-Russell (HR) diagram, with the two contours, from inside to outside, containing 68\% and 95\% probabilities, respectively. The addition of the UV filter greatly improves the parameter constraints compared to those achieved using only optical filters (see Fig.~3 of \citetalias{Maund2016}). Note, however, that this star appears in the Hertzsprung gap and is far off the MS. This also argues against it being a background star unrelated to SN~2006jc; such a possibility is very low since stars evolve very rapidly across the less populated Hertzsprung gap.

\subsubsection{Implications for the progenitor system}
\label{evidence06jc.sec}

If one assumes this star to have evolved without any binary interaction (i.e. as if it were a single star), it will have a mass of $M_2$~=~12.3$^{+2.3}_{-1.5}$~$M_\odot$ and an age of  $\tau$~=~17.4$^{+4.0}_{-3.9}$~Myr. This is obtained by comparing the star's position on the HR diagram with PARSEC (v1.2S) stellar isochrones \citep{Bressan2012}. In this case, the primary star (i.e. SN~2006jc's progenitor star) has a very similar mass of $M_1$~=~12.9$^{+2.5}_{-1.6}$~$M_\odot$ (we obtain this value by seeing, based on the PARSEC models, what mass of a single star would undergo core collapse at the derived age). However, these results are in severe contradiction to SN~2006jc's being a Type~Ibn SN. A primary star of this mass is unable to lose its envelope and evolve into a WR star solely via its stellar wind \citep{Crowther2007}. Instead, it will end up as a red supergiant (RSG), still retaining a massive hydrogen envelope, before it explodes as a hydrogen-rich Type~II-P SN \citep{Smartt2009}. Moreover, although a RSG can undergo significant mass loss, its CSM is usually not dense enough to produce the narrow spectral lines during the interaction with SN ejecta \citep[][the CSM would be hydrogen-rich as well]{Smith2017}. All these are inconsistent with observations.

Thus, the assumption made at the beginning of the previous paragraph is not correct; in other words, binary interaction must have occurred and significantly influenced the evolution of SN~2006jc's progenitor star and its companion. Precise modelling of their pre-SN evolution is beyond the scope of this paper. The evolution of the companion star depends on how much material it has accreted from the primary. If the companion star has not accreted much material at all (i.e. the mass transfer efficiency is very low, $\beta$~$\sim$~0), its evolution should have been much like that of a single star; in this case, its initial mass ($M_2$) and age ($\tau$) are still equal to the values as derived in the previous paragraph. On the other hand, if the mass transfer has been very efficient ($\beta$~$>$~0), the evolution of the companion in turn depends on when the mass transfer occurred (\citealt{Podsiadlowski1992}, see also Fig.~2 of \citealt{Claeys2011} for the response of the accreting star). It is possible that mass accretion occurred when the companion star was still on its MS; in this case, it became rejuvenated and then evolved like a single star of higher mass to its current position on the HR diagram. Alternatively, the mass accretion may occur when the companion star had already evolved off the MS; if so, it avoided rejuvenation but mass accretion was still able to shift its position on the HR diagram toward higher luminosities. Thus, with efficient mass transfer ($\beta$~$>$~0) the initial mass ($M_2$) and age ($\tau$) of the companion star should be smaller and larger than the values as derived in the previous paragraph, respectively. Considering all possible cases, we suggest an initial mass of $M_2$~$\leq$~12.3$^{+2.3}_{-1.5}$~$M_\odot$ and an age of $\tau$~$\geq$~17.4$^{+4.0}_{-3.9}$~Myr for the companion star.

It is not trivial to estimate the initial mass of the primary star (i.e. SN~2006jc's progenitor star), but we suggest that it should be very similar to that of the secondary. In Fig.~\ref{hrd06jc.fig} we show the locus of predicted binary companions on the HR diagram at the death of the primary star. The locus is predicted by BPASS \citep[binary population and spectral synthesis;][]{Eldridge2009}, which models binary systems with primary-to-secondary mass ratios of $q$~=~$M_2/M_1$~=~0.1 to 0.9 in steps of 0.1. It is apparent that, for most of the modelled systems, the secondary stars still reside on the MS and have much higher effective temperatures than that of SN~2006jc's binary companion. As the primary stars are more massive, they evolve on much shorter timescales than the secondary; a more evolved companion is expected only if the initial mass ratio is very close to unity \citep[see also][]{Zapartas2017}. Thus, for SN~2006jc's progenitor system, the two member stars should have very similar initial masses, such that the secondary could have enough time to evolve into the Hertzsprung gap before the primary star exploded as a SN.

Based on the above analysis, we suggest that SN~2006jc may not have arisen from a massive ($M_{\rm ZAMS}$~$>$~30~$M_\odot$) WR star via single-star evolution. Instead, its progenitor may be a much lower-mass star in an interacting binary system (with $q$~$\sim$~1.0, $M_2$~$\le$~12.3$^{+2.3}_{-1.5}$~$M_\odot$). The progenitor star's hydrogen envelope is most likely to have been stripped via binary interaction.

\subsubsection{More possibilities}
\label{pos06jc.sec}

\begin{figure}
\centering
\includegraphics[scale=0.8, angle=0]{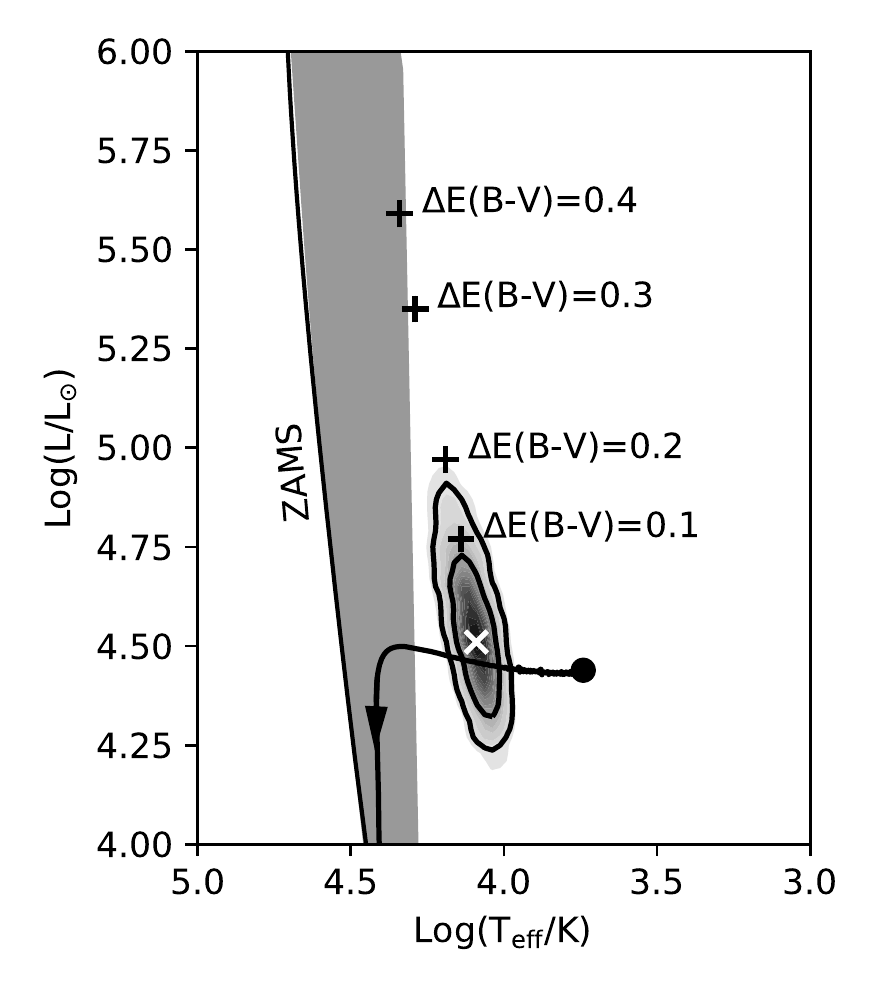}
\caption{`$\times$' symbol and contours: position of SN~2006jc's binary companion as derived in Section~\ref{companion06jc.sec} (same as those in Fig.~\ref{hrd06jc.fig}). Plus symbols: possible positions of SN~2006jc's binary companion, if it is subject to increased reddening caused by the newly formed dust. Solid curve: a possible evolutionary track (with the filled circle corresponding to the starting point) of a companion star after it has been inflated through interaction with the SN ejecta (see text for the adopted parameters). The grey shaded region along the ZAMS indicates the locus of binary companions at the death of the primary as predicted by the BPASS models.}
\label{pos06jc.fig}
\end{figure}

In the above analysis we have adopted a low interstellar reddening, which was obtained with SN~2006jc's spectrum at early times \citepalias[][supplementary material]{Pastorello2007}. SN~2006jc has, however, been observed to produce a significant amount of dust $\sim$70~days after explosion \citep{DiCarlo2008, Mattila2008, Nozawa2008, Smith2008, Sakon2009}. Thus, one cannot exclude the possibility that the binary companion may be subject to a higher reddening if it is obscured by the newly formed dust. In Fig.~\ref{pos06jc.fig} we show the companion star's positions on the HR diagram, assuming increased reddening values (plus symbols). These positions are derived by stellar SED fitting in the same way as before, except for increasing the reddening by $\Delta E(B-V)$~=~0.1, 0.2, 0.3, and 0.4~mag. In this scenario, the companion star could have a higher effective temperature and a higher luminosity. However, if the reddening is too high, the companion star (and the primary) would be too massive and inconsistent with the environment analysis in Section~\ref{env06jc.sec}. Measuring the stellar SED with higher precision may help to constrain the reddening and stellar parameters simultaneously.

The SN ejecta could inject energy into the companion star's surface layers when they collide with each other \citep{Hirai2014, Hirai2018, Hirai2015, Liu2015}. This ejecta-companion interaction (ECI) will quickly make the companion cooler and more luminous before it evolves back to the original state over a thermal timescale of the energy-injected layers, lasting years to decades. It is possible that SN~2006jc's companion star is a low-mass MS star which has been inflated in this way; by the time of the 2010 and 2017 observations, it was still bloated up by ECI and had not yet returned to its original state. As an example, the solid curve in Fig.~\ref{pos06jc.fig} shows a possible post-ECI evolutionary track for an ECI-inflated binary companion. In this model, the binary separation is 57~$R_\odot$, and the companion is a 4-$M_\odot$ MS star with an initial radius of 2.5~$R_\odot$; the primary explodes as a SN with an ejecta mass of 5~$M_\odot$ and an explosion energy of 10$^{52}$~ergs \citep[taken from estimates by][]{Tominaga2008}. The starting point of the track (filled circle in Fig.~\ref{pos06jc.fig}) marks the state of the companion star just after the ECI, when it was inflated to a lower effective temperature and a higher luminosity. The companion star was contracting back across the Hertzsprung gap when it was observed in 2010 and 2017, and eventually it will return to the MS and fade back to its original luminosity. The timescale over which this contraction happens strongly depends on the pre-SN binary separation and explosion parameters. In the above model, the companion star should have returned to the MS soon after the 2017 observations [with log($T_{\rm eff}$/K)~$\sim$~4.4]; its luminosity is now on the decline and will be around log($L/L_\odot$)~$\sim$~4.0 in the next few years. Thus, observations at a future epoch will be able to test such a scenario and to constrain the properties of the pre-SN binary configuration and the explosion itself.

If this is the case, the small secondary mass implies that the initial binary most likely experienced a common-envelope phase. Common-envelope phases are more efficient in getting rid of the entire hydrogen envelope than stable mass transfer (where stellar wind is still needed to shed the final bit of hydrogen). It will also naturally bring the binary separation closer such that ECI will have strong effects. Even in this case the mass of the primary star (i.e. the SN progenitor) can still be constrained, since the secondary-to-primary mass ratio must be large enough (or in other words, the binding energy of the primary star's envelope must be small enough) to allow a successful ejection of the whole envelope (otherwise, the secondary will merge with the core of the primary; see \citealt{Ivanova2013} for a review on common-envelope evolution). The small amount of low-velocity hydrogen detected at late times \citepalias[$\gtrsim$40~days after maximum;][]{Foley2007, Pastorello2008} could be attributed to mass stripped off the surface of the companion by the SN ejecta.

In summary, the possible effects of ECI and/or newly formed dust may complicate the analysis of SN~2006jc's binary companion. The existing observations are not able to confirm or reject such possibilities. If they do have a significant effect, the properties of the companion star (e.g. initial mass, $M_2$, and age, $\tau$) may be different from the results as derived in Section~\ref{evidence06jc.sec}, where we assumed no ECI and a low interstellar reddening. However, the conclusion reached in Section~\ref{evidence06jc.sec} should remain qualitatively robust that SN~2006jc was produced not by a massive WR star but a lower-mass progenitor in an interacting binary system. Such a conclusion is also supported by the environment analysis in Section~\ref{env06jc.sec}. The environment analysis is not affected by ECI or newly formed dust, since the stars surrounding SN~2006jc are at far distances from it (see Section~\ref{env06jc.sec} for details).

\subsection{SN2015G}
\label{companion15g.sec}

\begin{figure}
\centering
\includegraphics[scale=0.8, angle=0]{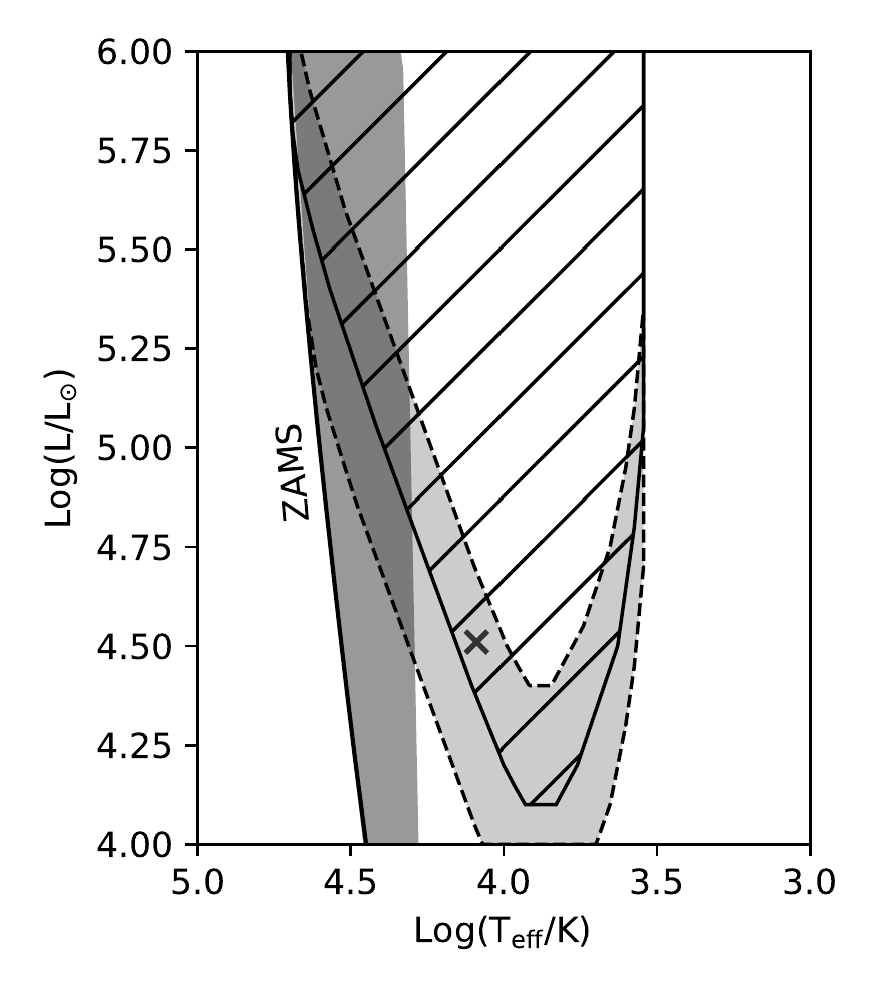}
\caption{Hatched region: forbidden area of SN~2015G's binary companion on the HR diagram. Shaded region enclosed by the dashed lines: uncertainty of the forbidden area by considering two extremes of distance and reddening values (see text). For comparison, the `$\times$' symbol indicates the most likely position for SN~2006jc's companion. The grey shaded region along the ZAMS indicates the locus of binary companions at the death of the primary as predicted by the BPASS models.}
\label{hrd15g.fig}
\end{figure}

\citetalias{Shivvers2017} have argued that SN~2015G's progenitor was not a massive WR star but a lower-mass one that lost its envelope through interaction with a binary companion. Given the non-detection of any late-time source at the SN position (Section~\ref{source15g.sec}), we attempt to constrain the properties of SN~2015G's binary companion, if present. For all possible positions on the HR diagram ($T_{\rm eff}$, $L$), we calculate the synthetic {\it WFC3/F300X} and {\it WFC3/F475X} magnitudes based on ATLAS9 \citep{Castelli2004} model spectra for single supergiants. A reddening of $E(B-V)$~=~0.384~mag and a distance of $D$~=~23.2~Mpc are used, both consistent with \citetalias{Shivvers2017}. We consider a  ($T_{\rm eff}$, $L$) position on the HR diagram to be ``forbidden" if either of the synthetic magnitudes becomes brighter than the detection limit in the corresponding band (Section~\ref{source15g.sec}). In this way, we derive a forbidden area for SN~2015G's binary companion on the HR diagram (the hatched region in Fig.~\ref{hrd15g.fig}).

Note that there is a large spread in the distance estimates for SN~2015G, ranging from 33.0 down to 16.2 Mpc \citep[e.g.][]{Vinko2001, Sorce2014}. For the dust reddening toward SN~2015G, \citetalias{Shivvers2017} estimated the host galaxy's contribution to be $E(B-V)$~=~0.053~$\pm$~0.028~mag using the sodium D1 line and $E(B-V)$~=~0.076~$\pm$~0.028~mag using the D2 line [the total reddening in the previous paragraph equals the average of these two values plus a Galactic contribution of $E(B-V)_{\rm MW}$~=~0.3189~mag]. In order to assess the effect of distance and reddening uncertainties, we further repeat the calculation of the forbidden region by considering two extremes. In the ``best" case, the smallest distance and lowest reddening are used while in the ``worst" case the largest distance and highest reddening are adopted. In Fig.~\ref{hrd15g.fig}, the dashed lines show the boundaries of the derived forbidden areas in these two extremes, which reflect the uncertainty of the forbidden area caused by distance and reddening errors.

Unfortunately, the detection limits cannot place very tight constraints on the companion star's properties. This is mainly because of the high extinction it suffers, especially in the UV band. We try to determine an upper limit for its {\it current} mass by comparing the derived forbidden area with BPASS products (the grey shaded region in Fig.~\ref{hrd15g.fig}) on the HR diagram. The upper mass limit depends on the companion star's effective temperature. If the companion star is still close to the ZAMS, its upper mass limit can reach as massive as 60$^{+35}_{-25}$~$M_\odot$. At the coolest end [log($T_{\rm eff}$/K)~$\sim$~4.3], however, the companion star's current mass cannot exceed 27$^{+6}_{-5}$~$M_\odot$. Interestingly, the derived forbidden area cannot exclude an even cooler companion star with low luminosities. We note that SN~2006jc's companion star lies just close to the boundary of the forbidden area on the HR diagram. Thus, it is possible that SNe~2006jc and 2015G may have similar companions and that their progenitor systems may have experienced similar pre-SN evolution with binary interaction. It is also possible that SN~2015G's companion is a compact object, such as a neutron star or a black hole.

\section{Environment}
\label{env.sec}

\begin{figure}
\centering
\includegraphics[scale=0.8, angle=0]{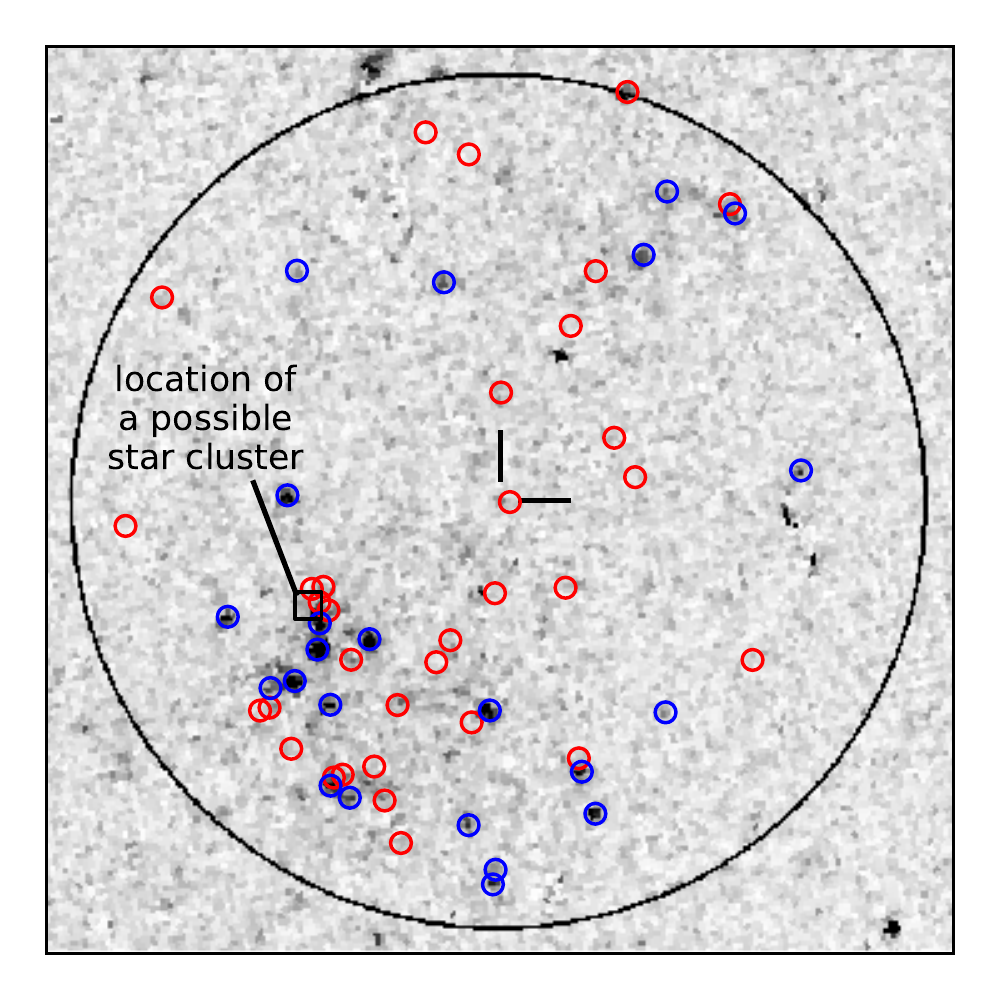}
\caption{Positions of detected stars (small circles) in the environment of SN~2006jc overlaid on the {\it WFC3/F300X} image. The blue circles corresponds to stars detected in the {\it WFC3/F300X} band, while the red ones are only detected in the optical bands. The black square indicates the location of a possible star cluster, which is undetected in the {\it WFC3/F300X} band. The large circle is centred on the SN site (shown by the crosshair) and has a radius of 400~pc. North is up and east is to the left.}
\label{env06jc_im.fig}
\end{figure}

\begin{figure*}
\centering
\includegraphics[scale=0.6, angle=0]{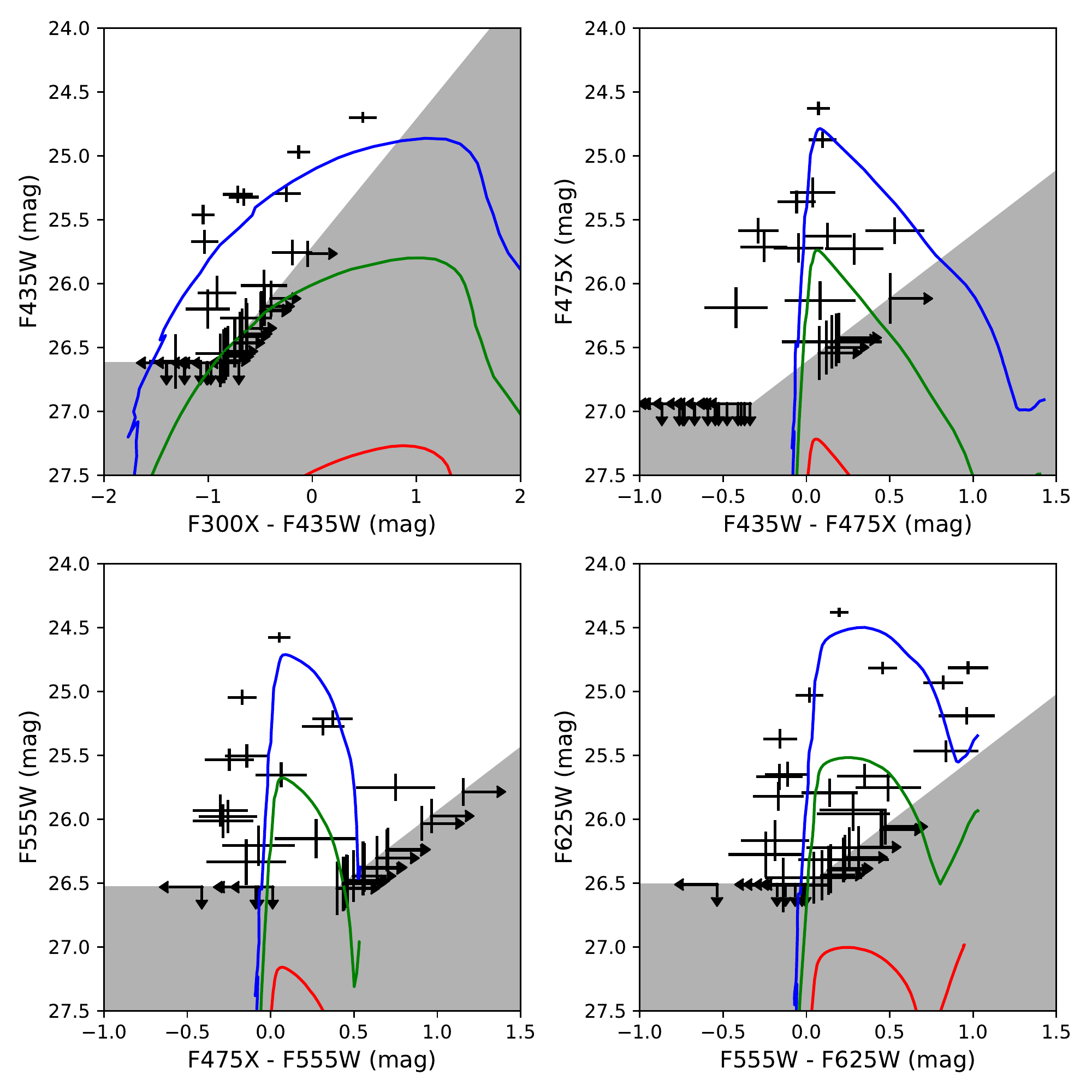}
\caption{Colour--magnitude diagrams of all stellar sources in the environment of SN~2006jc. Overlaid are single stellar isochrones corresponding to the three age components (blue: {\it Pop~A$_{\it 06jc}$}; green: {\it Pop~B$_{\it 06jc}$}; red: {\it Pop~C$_{\it 06jc}$}). See text for the values of their ages.}
\label{env06jc_cmd.fig}
\end{figure*}

\begin{figure}
\centering
\includegraphics[scale=0.8, angle=0]{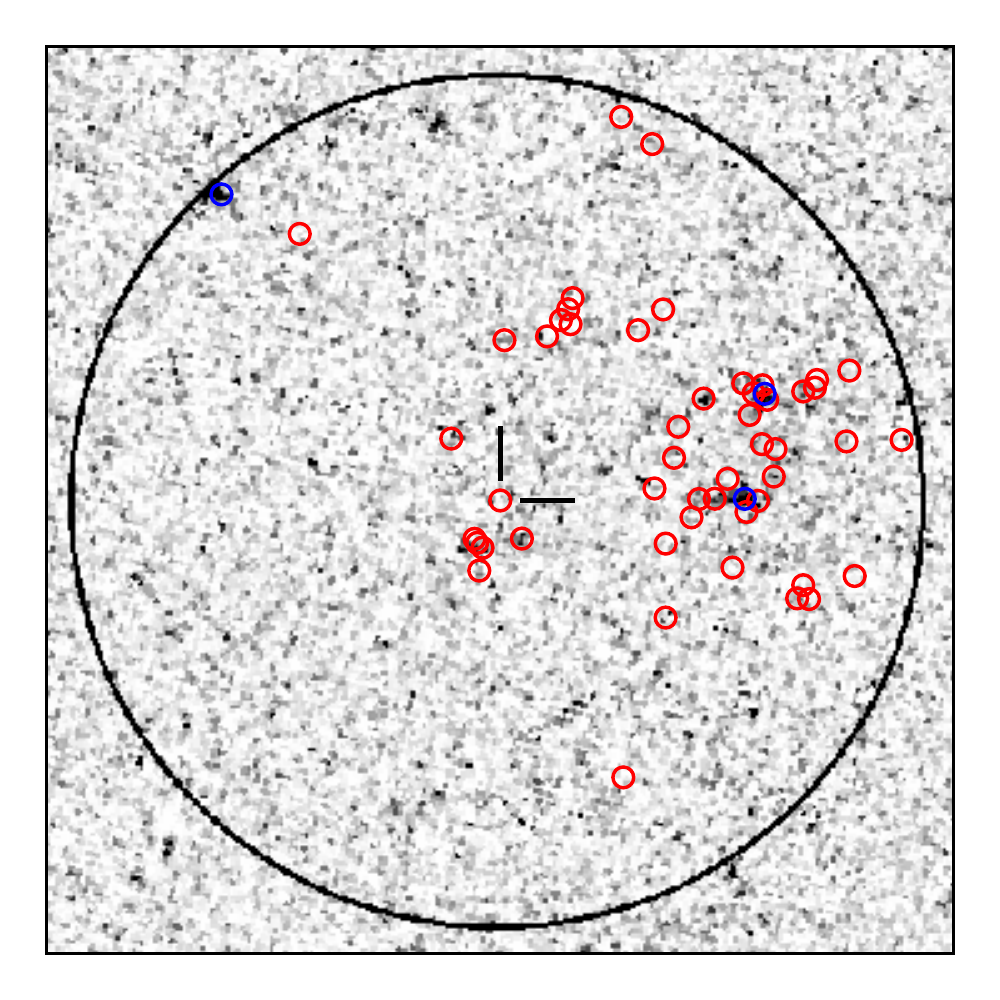}
\caption{Same as Fig.~\ref{env06jc_im.fig} but for SN~2015G.}
\label{env15g_im.fig}
\end{figure}

\begin{figure*}
\centering
\includegraphics[scale=0.6, angle=0]{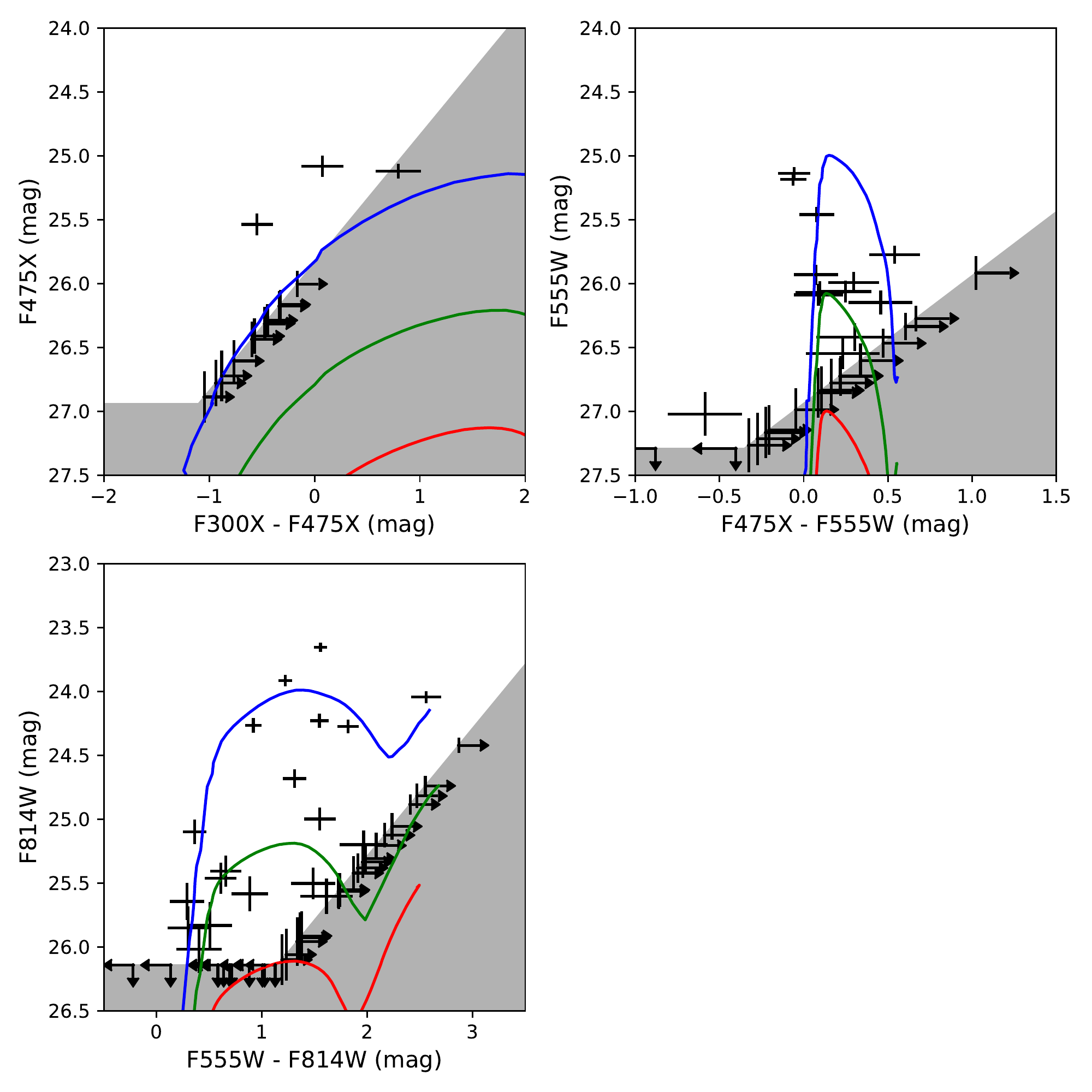}
\caption{Same as Fig.~\ref{env06jc_cmd.fig} but for SN~2015G.}
\label{env15g_cmd.fig}
\end{figure*}

\subsection{SN~2006jc}
\label{env06jc.sec}

Most massive stars are not born in isolation. Instead, they usually form in clusters \citep{Lada2003}, associations \citep[e.g.][]{Brown1994} or in groups that are hierarchically structured \citep[e.g.][]{Sun2017, Sun2017b, Sun2018}. During their short lifetimes, they can travel only a limited distance from their birthplaces before they explode as SNe. Thus, analysing the surrounding stellar populations provides important information on SN progenitors. In particular, the ages of their parental stellar populations correspond to the lifetimes of their progenitors; hence, they serve as a measure of the progenitors' initial masses \citep[e.g.][]{Maund2016, Maund2017, Maund2018}.

To quantitively analyse SN~2006jc's environment, we construct a stellar catalogue within 400~pc based on both the {\it ACS} and {\it WFC3} observations. In each band, we consider a detection as a reliable stellar source if its DOLPHOT quality parameters satisfy the following criteria:

(1) type of source, TYPE~=~1;

(2) photometry quality flag, FLAG~$\leq$~3;

(3) source crowding, CROWD~$<$~2;

(4) source sharpness, SHARP~$<$~0.5;

(5) signal-to-noise ratio, S/N~$\geq$~5.

\noindent
27, 29, 31, 23, and 20 stellar sources are reliably detected in the {\it ACS/F435W}, {\it ACS/F555W}, {\it ACS/F625W}, {\it WFC3/F300X}, and {\it WFC3/F475X} bands, respectively (In practice, we find that all sources with S/N~$\geq$~5 automatically meet the other criteria, suggesting that all these sources are point-like and have good photometry). They are then cross-matched with a searching radius of 1~pixel, and the final catalogue contains 57 stars in total. The positions of the stars are displayed in Fig.~\ref{env06jc_im.fig}. It is apparent that the SN is located in a relatively sparse area; it is in the vicinity of a clump of young, massive stars (to its southeast), but shows a clear projected offset from them by $\gtrsim$~200~pc. We also estimate a detection limit for each band using artificial stars randomly positioned in the circular area. 

Based on this catalogue, we apply a Bayesian approach (see \citealt{Maund2016b} for a detailed description) to determine the ages of SN~2006jc's surrounding stellar populations. In this analysis, we have assumed a \citet{Salpeter1955} initial mass function and a 50\% binary fraction; the binaries are considered as non-interacting systems with a flat primary-to-secondary mass ratio distribution between 0 and 1. The best-fitting result suggests that the interstellar reddening is very low [$E(B-V)$~=~0.09~mag, which is consistent with the estimate by \citetalias{Pastorello2007}] and that there are at least three age components ({\it Pops~A$_{\it 06jc}$}, {\it B$_{\it 06jc}$}, and {\it C$_{\it 06jc}$}, hereafter) of 10.2$_{-0.2}^{+0.2}$, 14.8$_{-0.3}^{+0.7}$, and 31.6$_{-0.7}^{+0.7}$~Myr, respectively. Figure~\ref{env06jc_cmd.fig} shows the colour-magnitude diagrams of the surrounding stellar populations, overlaid by single stellar isochrones corresponding to the derived ages.

These age estimates suggest that the Type~Ibn SN~2006jc is not likely to be produced by a massive WR star via single star evolution. In the single star channel, the minimum progenitor mass to become a WR star \citep[$M_{\rm ZAMS}$~$\geq$~30~$M_\odot$ at half-solar metallicity;][]{Crowther2007} corresponds to a stellar lifetime of only $\sim$6.4~Myr, which is significantly younger than {\it Pop~A$_{\it 06jc}$}, let alone the even older {\it Pop~B$_{\it 06jc}$} and {\it Pop~C$_{\it 06jc}$}. Any stars from these populations, which were massive enough to become WR stars, should have already exploded at least millions of years ago and cannot produce such a recent SN event.

We find a point source in the catalogue with much higher optical brightness than the other stars ($m_{F555W}$~=~24.01~$\pm$~0.03~mag). It also has significant H$\alpha$ emission with $m_{F658N}$~=~23.01~$\pm$~0.10~mag. Thus, it may be an unresolved star cluster (see Fig.~\ref{env06jc_im.fig} for its location) and has been excluded from the above population analysis. This source, however, is not significantly detected in the {\it WFC3/F300X} band. The lack of any UV emission suggests that it is not likely to host any young, massive stars which could evolve into WR stars. Apart from this possible star cluster, the {\it ACS/F658N} image does not show any significant H$\alpha$ emission in this area, suggesting very little star-forming gas. In summary, the analysis of the environment does not support a massive WR star as SN~2006jc's progenitor.

Recall that in Section~\ref{companion06jc.sec} we have derived an age estimate for SN~2006jc's progenitor as $\tau$~$\geq$~17.4$^{+4.0}_{-3.9}$~Myr (which depends on the amount of material the companion star has accreted from the primary). The lower age limit is significantly older than {\it Pop~A$_{\it 06jc}$} by 3.5$\sigma$, suggesting that the progenitor is not likely to be coeval with {\it Pop~A$_{\it 06jc}$}. On the other hand, the lower age limit is only slightly older than {\it Pop~B$_{\it 06jc}$} by 0.6$\sigma$. Thus, the SN progenitor may be coeval with either {\it Pop~B$_{\it 06jc}$} or {\it Pop~C$_{\it 06jc}$}, in which case the companion star has not or has accreted much material from the primary, respectively.


\subsection{SN~2015G}
\label{env15g.sec}

\citetalias{Shivvers2017} have analysed the surrounding stellar populations around SN~2015G based on the {\it WFC3/F555W} and {\it WFC3/F814W} photometry (i.e. Program~14149). They found that the highest initial mass the progenitor star could have is only 18~$M_\odot$, assuming that stars are coeval in this region and that the progenitor star has not been rejuvenated as a result of binary interaction. With the new observations in two more filters ({\it WFC3/F300X} and {\it WFC3/F475X}), we try to investigate the surrounding stellar populations more precisely.

To do this, we build a stellar catalogue within 400~pc from the SN site in the same way as for SN~2006jc. The catalogue contains a total of 53 stars, among which 32, 38, 3, and 15 stars are detected in the {\it WFC3/F555W}, {\it WFC3/F814W}, {\it WFC3/F300X}, and {\it WFC3/F475X} bands, respectively (all sources with S/N~$\geq$~5 are point-like and there is no evidence for any star clusters in this region). Their spatial distribution is shown in Fig.~\ref{env15g_im.fig}. As \citetalias{Shivvers2017} pointed out, SN~2015G has an environment which is very similar to that of SN~2006jc: both SNe occur in relatively sparse areas near clumps of young, massive stars but offset from them by $\sim$200~pc.

The ages of the surrounding populations are also estimated in the same way as for SN~2006jc. The best-fitting result suggests an interstellar reddening of $E(B-V)$~=~0.32~mag, which is consistent with the estimate by \citetalias{Shivvers2017}. We also find that there are at least three age components ({\it Pops~A$_{\it 15G}$}, {\it B$_{\it 15G}$}, and {\it C$_{\it 15G}$}, hereafter) of 9.8$_{-0.4}^{+0.5}$~Myr, 14.8$_{-0.3}^{+0.7}$~Myr, and 23.4$_{-1.6}^{+0.5}$~Myr, respectively. The colour-magnitude diagrams of the surrounding stellar populations are shown in Fig.~\ref{env15g_cmd.fig}, along with single stellar isochrones corresponding to the three age components. From this result we notice another striking similarity between the environments of SNe~2006jc and 2015G: the ages of {\it Pop~A$_{\it 15G}$} and {\it Pop~B$_{\it 15G}$} are consistent with those of {\it Pop~A$_{\it 06jc}$} and {\it Pop~B$_{\it 06jc}$} within errors, respectively. However, it is not clear whether this similarity is physically related to the SNe or is a mere coincidence.

As mentioned in the previous subsection, a single-star SN~Ibn progenitor is expected to have a very short lifetime of $\sim$6.4~Myr. This is much shorter than the ages of the three age components in the environment of SN~2015G. In other words, if the SN progenitor was a single star coeval with the surrounding stellar populations, the maximum initial mass it could have is only 18.8~$M_\odot$ (the lifetime of a star of this mass corresponds to the age of the youngest component). This value, which is very similar to the estimate by \citetalias{Shivvers2017}, is much smaller than the initial mass of WR stars \citep[$M_{\rm ZAMS}$~$\ge$~30~$M_\odot$ at half-solar metallicity;][]{Crowther2007}. Thus, this analysis supports \citetalias{Shivvers2017}'s conclusion that SN~2015G was produced in an interacting binary system.


\section{Discussion}
\label{discussion.sec}

\subsection{Progenitor channels of SNe~Ibn}
\label{channel.sec}

As mentioned in Section~\ref{intro.sec}, there are expected to be two main progenitor channels for stripped-envelope SNe: single, massive WR stars \citep{Gaskell1986} and lower-mass stars in interacting binaries \citep{Podsiadlowski1992}. Observations suggest that, for the "normal" stripped-envelope SNe (i.e. IIb, Ib, Ic), the binary channel seems to be dominant. For example, statistical studies show that their occurrence rate is too high to be reconciled solely by the single-star channel  \citep{Smartt2009b, Li2011, Smith2011, Shivvers2017b, Graur2017}. Meanwhile, light curve modelling suggests that they have low ejecta masses ($\sim$3~$M_\odot$), which are consistent with binary models \citep{Drout2011, Taddia2015, Lyman2016}. Furthermore, a series of progenitor non-detections are also compatible with the low luminosities of lower-mass progenitors in interacting binaries \citep[e.g.][]{Maund2005, Maund2005b, Eldridge2013}.

Early work on SNe~Ibn has assumed that they come from massive WR stars via single star evolution \citepalias[e.g.][]{Foley2007, Pastorello2007, Pastorello2008}. This argument is supported by their CSM velocities \citep[$\sim$1000--2000~km~s$^{-1}$ as derived from the narrow helium lines;][]{Pastorello2016}, which are comparable to the typical wind velocity of WR stars \citep{Crowther2007}\footnote{Some transitional SNe~Ibn (e.g. SNe~2005la, 2011hw), which show narrow lines of both hydrogen and helium, may have much lower CSM velocities ($\lesssim$500~km~s$^{-1}$). They have been suggested to arise from stars transitioning between the LBV and the WR stages \citep[e.g.][]{Pastorello2008b, Pastorello2015}.}. This may, however, be a coincidence and cannot rule out other types of progenitors. The SN~Ibn OGLE-2014-SN-131 \citep[OGLE14-131 hereafter;][]{Karamehmetoglu2017} has a long rise time and a broad light curve, which indicates a large ejecta mass of $M_{\rm ej}$~=~18~$M_\odot$. This suggests that OGLE14-131 may indeed have a massive WR progenitor (see e.g. Fig.~10 of \citealt{Lyman2016} for the ejecta mass distributions of different progenitor channels). Most SNe~Ibn, however, have very narrow and fast evolving light curves \citep{Hosseinzadeh2017}. For example, the well-observed Type~Ibn SN 2018bcc has a very small ejecta mass of $M_{\rm ej}$~=~1.3--1.8~$M_\odot$ \citep{Karamehmetoglu2019}, which may point towards a much lower-mass progenitor (here we caution that the ejecta masses and other explosion parameters from light curve modelling are only indicative as they may be subject to degeneracies or systematic uncertainties). As no SN~Ibn progenitors have been detected in pre-explosion images, their connection to WR stars is by no means conclusive.

The analysis in the previous sections suggests that SNe~Ibn may also form in lower-mass, interacting binary systems. It is quite surprising that even the class prototype SN~2006jc is produced through this channel. This raises the question of whether interacting binaries are a special or a common progenitor channel for the SNe~Ibn population. Just in terms of light curves, SNe~2006jc and 2015G are not special at all and the light curves of SNe~Ibn are very homogeneous \citep[despite a few exceptions;][]{Hosseinzadeh2017}. If this homogeneity reflects a uniformity in their progenitors, most SNe~Ibn should be produced in interacting binary systems just like SNe~2006jc and 2015G. However, this does require further investigation and we note that the light curve modelling of SNe~Ibn are subject to significant uncertainties \citep[e.g.][]{Tominaga2008, Chugai2009, Moriya2016}.

\citet{Hosseinzadeh2019} have analysed the environments of SNe~Ibn. They found that, when excluding the peculiar PS1-12sk, the distribution of local UV surface brightness is not significantly different from those for other types of core-collapse SNe (their Fig.~3). Thus, most SNe~Ibn may have progenitors of similar ages to those of normal stripped-envelope SNe (which, as mentioned before, are thought to be mainly produced in the binary channel). This may be further evidence that SNe~Ibn are produced predominantly via the binary channel. We do, however, caution the risk that low-resolution images may not reflect the precise relation between the SN progenitors and their environments \citep[e.g. see the discussion of][]{Maund2016b}.

\subsection{Pre-SN giant eruptions of stripped-envelope stars}
\label{eruption.sec}

\subsubsection{CSM and pre-SN mass loss of interacting SNe}
\label{general.sec}

SN~2006jc exhibited narrow helium lines in its spectra (with full widths at half maximum of $\sim$2000~km~s$^{-1}$; \citealt{Anupama2009}; \citetalias{Foley2007, Pastorello2007}) and a unique rise in the X-ray light curve over $\sim$4~months after the explosion \citep{Immler2008}. Both signatures indicate a strong interaction between the SN ejecta and CSM abound the progenitor. Detailed analysis shows that the CSM is located at a distance of $\sim$10$^{15}$--10$^{16}$~cm from the SN progenitor and is a few times 0.01~$M_\odot$ \citep{Anupama2009, Immler2008}. This corresponds to a mass-loss rate of $\sim$10$^{-1}$~$M_\odot$~yr$^{-1}$, assuming a timescale of 0.1~year for the mass-loss event. The narrow helium lines, in particular, suggest the CSM to be very dense, such that it can slow the forward shock and form a cool, dense shell (CDS) via radiative cooling at the ejecta-CSM interface \citep{Chugai2004}. Such a dense CSM cannot be formed via relatively steady winds from evolved stars, which do not reach the required mass-loss rate \citep{Smith2017}. Instead, it is most likely to have formed from a giant eruption with a very high mass-loss rate, which occurred just before the SN and lasted for a very short timescale. An eruption for SN~2006jc was detected as an optical transient in 2004 \citepalias[UGC~4904-V1;][]{Pastorello2007}, which reached a peak magnitude of $M_R$~$<$~$-$14.1~mag and remained visible for a few days \citepalias{Pastorello2007}. Similarly, the CSM of SN~2015G was also formed in an eruptive mass-loss episode, which occurred just $\sim$1~year before the SN \citepalias{Shivvers2017}. Unfortunately, its pre-SN eruption was not detected in a 20~year's monitoring dataset \citepalias[][]{Shivvers2017} and it might have occurred during the gaps in the observations.

SNe with pre-SN eruptions (and thus strong ejecta-CSM interaction) suggest that massive stars may become wildly unstable and undergo extreme mass loss at the latest evolutionary stages. Such a phenomenon can significantly affect their evolution before core collapse, but has not been included in the standard stellar evolutionary models. Apart from SN~2006jc, pre-SN eruptions have also been directly observed for a few cases, such as SN~2009ip \citep{Mauerhan2013}, SN~2010mc \citep{Ofek2013}, LSQ13zm \citep{Tartaglia2016}, and SN~2015bh \citep{Elias-Rosa2016, Thone2017}. They all belong to Type~IIn, for which pre-SN outbursts seem to be quite common \citep{Ofek2014}. Direct progenitor detections have confirmed that SNe~IIn could be the terminal explosion of LBVs (e.g. SN~2005gl, \citealt{GalYam2007, GalYam2009}; SN~2009ip, \citealt{Mauerhan2013}; SN~2010jl, \citealt{Smith2011b}). Note, however, that the class of Type~IIn is very heterogeneous and some SNe~IIn may have other types of progenitors \citep[e.g. the long-lived SNe~1988Z and 2005ip may come from massive RSGs;][]{Smith2017b}.

LBVs have long been known to be massive stars with dramatic instability. Their initial masses are larger than $M_{\rm ZAMS}$~=~25~$M_\odot$ \citep{Smith2004, Vink2012}\footnote{Note, however, that their {\it final} masses can be as low as $\sim$10--15$M_\odot$ due to the extreme mass loss \citep{Vink2012}.} and the most extreme ones can reach $\gtrsim$100~$M_\odot$ \citep[e.g. $\eta$~Carina;][]{Kashi2010}. They exhibit a wide range of irregular variable phenomena, the most pronounced of which is their so-called giant eruptions. During the giant eruptions, they increase their bolometric luminosities for months to years accompanied by extreme mass loss \citep{Smith2014c}. Famous examples include $\eta$~Carina's {\it Giant Eruption} in the 19th century \citep{Smith2011c} and P~Cygni's 1600 AD eruption \citep{Smith2006}. Some ``SN impostors" may be the giant eruptions of extragalactic LBVs \citep[e.g.][]{VanDyk2012}. Although LBVs were believed to be a transitional phase between the MS and the WR stages \citep{Conti1976, Massey2003}, it is now clear that they can undergo core collapse as the direct progenitors of SNe~IIn.

\subsubsection{Which star was responsible for the pre-SN giant eruption?}
\label{which.sec}

The H-poor interacting SNe~Ibn raise a puzzling question about the pre-SN  giant eruption of their stripped-envelope progenitors. Early SNe~Ibn studies (when their progenitors were still believed to be massive WR stars) speculated that WR stars may still have some residual LBV-like instability and can produce pre-SN eruptions like that of SN~2006jc \citepalias{Foley2007, Pastorello2007, Pastorello2008}. However, this scenario is beyond our standard understanding of stellar evolution. To overcome this problem, an alternative scenario was proposed \citepalias{Pastorello2007, Pastorello2008}, invoking a binary system in which an LBV produced the eruption in 2004 and a WR star happened to explode two years later as SN~2006jc.

Our work suggests that the binary companion is not likely to be an LBV or to be responsible for the 2004 eruption (and the formation of the CSM). This argument is based on the following reasons: (1) the companion star has a much lower luminosity [log($L/L_\odot$)~=~4.52$^{+0.13}_{-0.13}$] than the typical luminosities [log($L/L_\odot$)~$\sim$~5.0--6.5] of known LBVs \citep{Smith2004, Smith2019}\footnote{The results of \citetalias{Maund2016} also argue against the companion star as an LBV, although its properties were not as tightly constrained.}; (2) the ACS/{\it F658N} non-detection, as \citetalias{Maund2016} pointed out, does not agree with the strong H$\alpha$ emission of LBVs; (3) all known LBVs with giant eruptions still retain their massive hydrogen-rich envelopes, which are inconsistent with the hydrogen-poor and helium-rich CSM of SN~2006jc; (4) the environment analysis in Section~\ref{env06jc.sec} argues against any very massive stars in this area; and (5) it is hard to explain the time synchronisation between the eruption in 2004 and the SN explosion in 2006, if they are physically unrelated to each other (thus, any possible mechanism for pre-SN eruptions should be able to explain this time synchronisation; see Section~\ref{mechanism.sec}). In summary, the companion star is most likely a normal supergiant, and the pre-SN giant eruption (and the CSM) should have been produced by SN~2006jc's progenitor star itself.

For SN~2015G, its pre-SN eruption should also have been produced by its progenitor star itself (note that its pre-SN eruption, as mentioned in Section~\ref{general.sec}, was not directly detected but inferred from the behaviour of the SN). The reasons are similar to those for SN~2006jc from its helium-rich CSM, environment, and the time synchronisation between the pre-SN eruption and SN explosion.

\subsubsection{Pre-SN giant eruption in lower-mass massive stars}
\label{lowmass.sec}

\begin{figure*}
\centering
\includegraphics[scale=0.22, angle=0]{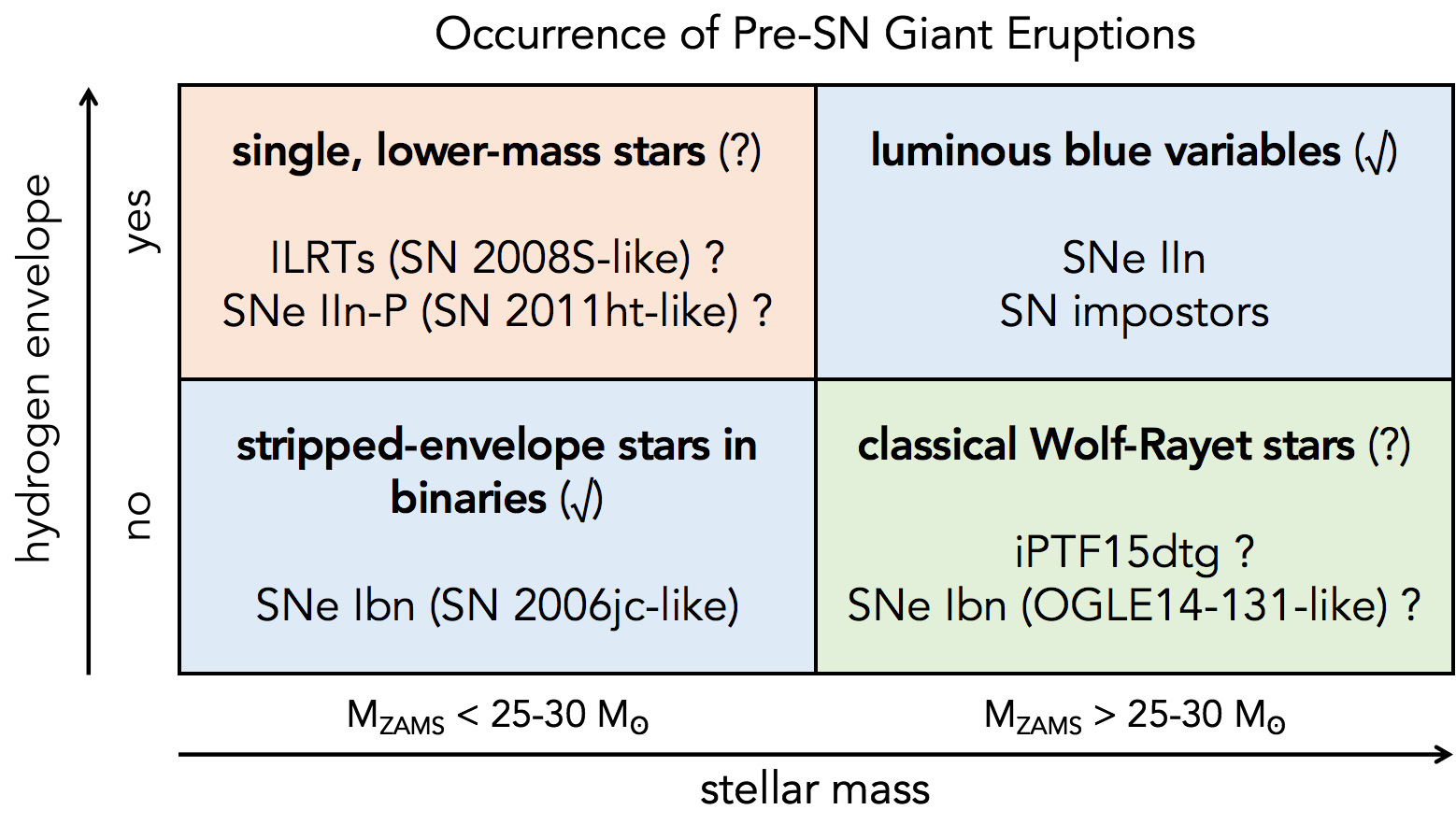}
\caption{Confirmed ($\checkmark$) and uncertain (?) possibilities of pre-SN giant eruptions in stars with different initial masses and with the presence/absence of hydrogen envelopes. We also list the phenomena which could imply the pre-SN giant eruptions for the corresponding types of stars. Phenomena whose implications are still uncertain are followed by question marks (see text).}
\label{summary.fig}
\end{figure*}

If SNe~2006jc and 2015G were produced via the binary channel (Sections~\ref{companion.sec} and \ref{env.sec}), the above discussion leads to a conclusion that {\it pre-SN giant eruptions can occur not only in massive LBVs but also in much lower-mass, stripped-envelope stars from interacting binary systems}. Figure~\ref{summary.fig} summarises the possibility of pre-SN giant eruptions in different types of stars. Note that we have borrowed the concept of ``giant eruptions" from LBVs to describe the mass-loss phenomena that have extreme mass-loss rates ($\dot{M}$~$\gtrsim$~10$^{-2}$~$M_\odot$~yr$^{-1}$), last for a short period (months to years, or even shorter) and are usually associated with optical brightening of several magnitudes. They do not necessarily arise from the same mechanism as for LBV giant eruptions, and we distinguish them from the other types of pre-SN mass loss that are less extreme. For example, (massive) RSGs may have superwinds with enhanced mass-loss rate before core collapse \citep{Heger1997}; compared with giant eruptions, however, the superwinds can blow for a much longer time (up to millennia) and their mass-loss rates are lower ($\dot{M}$~$\lesssim$~10$^{-3}$~$M_\odot$~yr$^{-1}$).

The possibility of LBV-like giant eruptions in ``lower-mass" massive stars has been discussed based on the study of SN~2008S-like events \citep[i.e. ILRTs, with NGC~300-OT being another prototype;][]{Prieto2008, Bond2009, Thompson2009}. Such events exhibit narrow Balmer and [Ca~{\footnotesize II}] lines in their spectra and, with peak magnitudes of $-$15~$\lesssim$~$M_V$~$\lesssim$~$-$13, are much fainter than normal core-collapse SNe. Their progenitors are deeply dust-enshrouded stars with extremely red mid-infrared colours and relatively low bolometric luminosities ($\sim$5~$\times$~10$^4$~$L_\odot$). Analysis shows that their progenitors have much lower initial masses \citep[e.g. $\sim$10~$M_\odot$ for SN~2008S, while a range of values between 10 and 25~$M_\odot$ has been reported for NGC~300-OT;][]{Prieto2008, Bond2009, Berger2009, Gogarten2009} than those of typical LBVs ($M_{\rm ZAMS}$~$\ge$~25~$M_\odot$). \citet{Smith2009b}, \citet{Bond2009}, and \citet{Berger2009} argued that these events are not real SN explosions but the giant eruptions of their progenitor stars. If so, the eruptive phenomena seen in LBVs can extend down to much lower-mass stars than previously thought. However, the nature of SN~2008S-like events is still under debate, and \citet{Thompson2009} and \citet{Botticella2009} suggested that they could be low-energy, electron-capture SNe from super-AGB stars. Furthermore, even if such events are indeed giant eruptions of lower-mass stars, it is not yet clear whether they occur just before the core collapse or in the middle of the lifetimes of their progenitors (i.e. whether they are {\it pre-SN} or not). Thus, more observations are still needed to confirm the nature of SN~2008-like events.

SNe~IIn-P are those who exhibit narrow hydrogen lines like SNe~IIn but have plateau light curves similar to SNe~II-P (e.g. SN~1994W, SN~2009kn, SN~2011ht, \citealt{Mauerhan2013b}). The \citet{Chugai2004b} model suggests that SN~1994W had a very high pre-SN mass-loss rate of $\dot{M}$~=~0.3~$M_\odot$~yr$^{-1}$ at $\sim$1.5~year before core collapse, and a precursor outburst has been detected one year prior to the explosion of SN~2011ht \citep{Fraser2013}. These SNe have very low $^{56}$Ni masses ($\leq$~0.0026~$M_\odot$ for SN~1994W, \citealt{Sollerman1998}; $<$~0.023~$M_\odot$ for SN~2009kn, \citealt{Kankare2012}; 0.006--0.01~$M_\odot$ for SN~2011ht, \citealt{Mauerhan2013b}), which are consistent with electron-capture SNe from the lowest-mass ($M_{\rm ZAMS}$~$\sim$~8--10~$M_\odot$) stars that can undergo core collapse. If so, pre-SN giant eruptions may also occur in these much lower-mass, hydrogen-rich stars. However, the low $^{56}$Ni masses can also result from massive ($M_{\rm ZAMS}$~$>$~25--30~$M_\odot$) progenitors if a substantial fraction of the inner core ejecta falls back on to the compact remnant \citep{Fryer1999}. Furthurmore, the non-detection of SN~2011ht's progenitor cannot rule out LBVs in their quiescent phase \citep{Roming2012}. Thus, it is still not fully clear what types of progenitors can give rise to SNe~IIn-P. As a result, we leave the implications of SNe~IIn-P as uncertain in Fig.~\ref{summary.fig}.

Early SN spectra taken within days of explosion (flash spectroscopy) have revealed narrow emission lines of high-ionisation species (e.g. SN~2016bkv, \citealt{Hosseinzadeh2018}; SN~1998S, \citealt{Shivvers2015}; SN~2013fs, \citealt{Yaron2017}). Such features arise from recombination of the CSM ionised by the shock-breakout flash; thus, flash spectroscopy can trace the CSM recently ejected from the progenitors. With this method, \citet{Khazov2016} shows that CSM are quite common among SNe~II. Many SNe with flash-ionised features may not come from massive LBV progenitors but still have significant pre-SN mass-loss rates reaching $\dot{M}$~$\sim$~10$^{-3}$~$M_\odot$~yr$^{-1}$ (e.g. SN~2013fs, \citealt{Yaron2017}). However, such mass-loss rates are not as extreme as those of the LBV-like giant eruptions ($\dot{M}$~$\gtrsim$~10$^{-2}$~$M_\odot$~yr$^{-1}$). Thus, their CSM is more likely to come from RSG superwinds or weaker outbursts of mass loss. Yet, we cannot exclude the possibility that flash spectroscopy may find LBV-like giant eruptions in ``lower-mass" massive stars with more future observations.

SNe~Ibn provide another opportunity to address this issue. Those produced via the binary channel serve as compelling evidence that pre-SN giant eruptions can indeed occur in their lower-mass, stripped-envelope progenitor stars. Combined with LBVs and SNe~IIn, they trace pre-SN giant eruptions in stars over a wider mass spectrum and with the presence/absence of hydrogen envelopes. Since SN~Ibn light curves decline much faster than those for SNe~IIn \citep{Hosseinzadeh2017}, \citet{Moriya2016} suggested that their pre-SN eruptions last for much shorter timescales and have higher mass-loss rates/CSM velocities compared to the long sustained mass loss in SN~IIn progenitors. This may reflect a difference in the stellar structures and/or in the energy sources for their eruptions. Single stars of similar initial masses still retain massive hydrogen envelopes before core collapse, and from our current understanding they may not be able to produce any LBV-like eruptions (note however the uncertain implications from SN~2008S-like events and SNe~IIn-P). This possibly suggests that the removal of hydrogen envelopes may aid the occurrence of giant eruptions in SN~Ibn progenitors, possibly because hydrogen stripping renders the luminosity-to-mass ratio ($L$/$M$) closer to the Eddington limit and thus makes the stars less stable.

On the other hand, it needs to be re-investigated whether WR stars can undergo pre-SN giant eruptions, since it becomes increasingly unclear whether they can be SN~Ibn progenitors. OGLE14-131, as mentioned in Section~\ref{channel.sec}, had a very long rise time and a broad light curve. \citet{Karamehmetoglu2017} modelled its light curve with a large ejecta mass of $M_{\rm ej}$~=~18~$M_\odot$, consistent with a massive WR progenitor. As cautioned by the authors, however, the light curve modelling may be subject to parameter degeneracy and systematic uncertainties especially if it is powered by ejecta-CSM interaction.

Some SNe of other types have also provided tentative evidence for pre-SN giant eruptions in WR stars. Using flash spectroscopy, \citet{GalYam2014} showed that the Type~IIb SN~2013b had a CSM very similar to WR winds. They suggested that it had a WR progenitor which experienced enhanced mass loss at $\dot{M}$~$>$~0.03~$M_\odot$~yr$^{-1}$ just one year before explosion. More quantitative analysis \citep{Groh2014}, however, seems to rule out a WR progenitor and favour an LBV or yellow hypergiant progenitor. \citet{Taddia2016} analysed the SN~Ic iPTF15dtg, which has two peaks in the light curve. Hydrodynamical modelling suggests that it has a large ejecta mass ($M_{\rm ej}$~$\sim$~10~$M_\odot$), implying a massive WR progenitor. The early peak can be reproduced if the progenitor was surrounded by an extended ($\gtrsim$~500~$R_\odot$), low-mass ($\gtrsim$~0.045~$M_\odot$) envelope. This envelope may have formed via an eruption from the progenitor at $\sim$~10~days before core collapse with an extreme mass-loss rate ($\dot{M}$~=~1.9~$M_\odot$~yr$^{-1}$). However, the envelope can also be explained by unstable mass transfer between the progenitor and its binary companion if the progenitor exploded during a common-envelope phase. Thus, more work is still needed to confirm whether WR stars can undergo the giant eruptions that are commonly seen in LBVs.

There are recent reports of precursor outbursts for ``normal" stripped-envelope SNe, which do not exhibit narrow spectral lines like SNe~IIn or Ibn. \citet{Corsi2014} found tentative evidence for a precursor outburst for PTF11qcj, and \citet{Ho2019} report a more definitive detection for 2018gep (i.e. ZTF18abukavn). Both SNe are broad-lined Type~Ic. For SN~2018gep, the mass loss rate, inferred by the early light curve, reaches a very high value of $\dot{M}$~=~0.6~$M_\odot$~yr$^{-1}$ in the days before SN explosion. The lack of narrow spectral lines are possibly because the CSM, which the shock runs through after the initial breakout, are of much lower density. Still, we do not fully understand whether their progenitors are massive WR stars or lower-mass stars from interacting binaries. With the advent of high-cadence transient surveys, we may be able to find more cases of pre-SN giant eruptions and get a more complete understanding of this intriguing phenomenon.

\subsubsection{Mass loss mechanisms}
\label{mechanism.sec}

As previously mentioned, stellar wind cannot reach the necessary mass-loss rate to form the dense CSM around SN~Ibn progenitors. The progenitor of SN~2006jc, for example, has a pre-SN mass-loss rate of $\sim$10$^{-1}$~$M_\odot$~yr$^{-1}$ \citep{Smith2017}; in contrast, the wind mass-loss rates of evolved stars (e.g. WR stars, RSGs, AGB stars) are typically not more than 10$^{-4}$~$M_\odot$~yr$^{-1}$ \citep{Smith2014c}. In this subsection, we review a few other mass-loss mechanisms and discuss whether they are applicable for SNe~2006jc/2015G's progenitors. We find that none of them fit the bill in an obvious way, and more efforts are still needed to understand what drives the pre-SN eruptions.

\paragraph*{S~Doradus-type mass loss}
S~Doradus-type outbursts, commonly seen for LBVs, are visual brightening that occurs when the peak of a star's SED shifts from the UV to visual wavelengths \citep{Vink2012}. This is caused by an iron opacity bump in the outer layers, which makes the star locally super-Eddington \citep{Grafener2012}. However, this mechanism cannot drive major mass-loss events. The typical mass-loss rate of an S~Doradus-type outburst is $\sim$10$^{-4.5}$~$M_\odot$~yr$^{-1}$, which is significantly smaller than those of most SN~Ibn/IIn progenitors ($>$10$^{-2}$~$M_\odot$~yr$^{-1}$) \citep{deKoter1996, Smith2014c}. Also it is unclear whether this mechanism could take place in H-poor stars such as SN~Ibn progenitors.

\paragraph*{Dynamical ejection from unstable (common) envelopes}
Some evolved stars (e.g. Mira variables and RSGs) may have dynamically unstable envelopes which develop large-amplitude pulsations that grow with time (e.g. \citealt{Wood1974}; \citealt{Yoon2010}; {\color{blue} Clayton et al. in preparation}). This instability has also been found in the numerical simulations of giant common envelopes of interacting binaries during the late spiral-in phase \citep{Clayton2017}. It has been proposed that dynamical mass ejection may occur episodically from such unstable envelopes. For example, \citet{Clayton2017} showed that the rebound following a high-amplitude compression, if it has not been damped by the non-linear effects of catastrophic cooling and internal decoherence, can be strong enough to accelerate a layer of matter at the stellar surface to above escape velocity.

Low-mass (below $\sim$3~$M_\odot$) helium stars, which are exclusively produced in binary systems, are potential candidates for SN~Ibn progenitors. When they evolve into helium giants (with radii of a few 100~$R_\odot$), their envelopes may become dynamically unstable and undergo episodic mass ejections. If they explode in this phase, they would presumably produce SNe Ibn as the SN ejecta interact with the previously ejected material. However, the typical ejection velocities are probably not more than 100--200~km~s$^{-1}$ and are significantly smaller than the CSM velocities of SNe~Ibn \citep[$\sim$1000--2000~km~s$^{-1}$;][]{Pastorello2016}.

Helium stars with masses of $\sim$3--6~$M_\odot$ also expand after core helium burning; they will experience a late mass-transfer phase if they have a close binary companion \citep{Tauris2015}. Unstable mass transfer may develop an extended common envelope around the system, in particular if the binary companion is of lower mass. The dynamical instability of the common envelope may lead to episodic dynamical mass ejections, which increase toward the epoch of supernova. For SN~2006jc, however, the observed companion was not able to trigger such a late common-envelope phase, since its initial mass was very close to that of the progenitor. This mechanism may be applicable to SN~2015G, if its progenitor had a low-mass close binary companion.

\paragraph*{Binary mergers}
Binary mergers are the inevitable consequence of common-envelope systems if the orbital energy released in the spiral-in phase is not sufficient to eject the envelope. Some LBV giant eruptions, such as that of $\eta$~Carinae, are possibly triggered by binary mergers (e.g., \citealt{Podsiadlowski2006}, \citealt{Justham2014}, {\color{blue} Hirai et al. in preparation}). \citet{Chevalier2012} proposed that SNe~Ibn may be produced via the merger of a helium star and an compact object (neutron star or black hole). In this scenario, the mass loss is driven by common-envelope evolution and the SN is triggered by the in-spiral of the compact object to the central core of the helium star. If SN~2006jc was produced in this way it must have been in a triple system, since it has a companion star which is still observable.

\paragraph*{Convection-excited waves}
Hydrodynamic waves may be excited by vigorous core convection and propagate outward at late nuclear burning stages \citep{Quataert2012, Shiode2014, Smith2014, Fuller2017, Fuller2018}. From core carbon burning and beyond, core temperatures and densities are sufficiently high that neutrino cooling dominates over other cooling processes. Neutrino cooling is very sensitive to the temperature, but nuclear burning has an even higher sensitivity. This difference leads to a large temperature gradient that drives vigorous convection in the core region. Simulations \citep[e.g.][]{Meakin2006, Meakin2007} show that the convection may excite hydrodynamic waves at the interface between the convective and non-convective zones. The waves propagate outward with a super-Eddington energy flux before they are dissipated in the outer regions. An outburst of mass loss could be triggered if the waves can reach close enough to the star's surface. This mechanism also nicely explains the time synchronisation between the outburst and the SN explosion, since the late nuclear burning stages last for very short timescales.

Following this idea, the calculation of \citet{Shiode2014} was successful in reproducing the ejecta mass, velocity, and energetics for SN~2006jc's pre-SN eruption. However, they have assumed a WR-star progenitor, which was found to be unable to produce any eruption earlier than $\sim$6 months before core collapse. For SN~2006jc, this timescale is 2 years and would require a very low helium core mass ($\sim$5~$M_\odot$) according to their calculation. We note that this condition is easily satisfied for a lower-mass progenitor star in an interacting binary system.

\citet{Fuller2018} also calculated this process in stars composed of a 5-$M_\odot$ helium core evolved from a 15-$M_\odot$ progenitor stripped of its hydrogen envelope. They found that wave heating can drive pre-SN eruptions with timescales and mass-loss rates consistent with observations. Yet, the calculated eruption luminosities are much smaller than that for SN~2006jc. The authors suggest that its progenitor may be more massive than their models, and that shell-shell collisions in the wave-driven wind may increase the luminosities. Interestingly, \citet{Fuller2017} showed that the convection-excited waves cause only mild pre-SN outburst in a 15-$M_\odot$ star, if its hydrogen envelope is not stripped. In this case, the waves thermalise their energy just outside the helium core, and the massive hydrogen envelope prevents most energy from diffusing outward. This may explain why pre-SN giant eruptions favour stripped-envelope stars in binaries instead of single stars of similarly low initial masses (see the discussion in Section~\ref{lowmass.sec}; note, however, the uncertain implications by SN~2008S-like events and SNe~IIn-P).

\section{Summary and Conclusions}
\label{summary.sec}

In this paper we report new UV and optical {\it HST} observations of two nearby Type~Ibn SNe~2006jc and 2015G. The observations were conducted at late times, when the SNe themselves have faded significantly. Combined with archival optical observations, we focus on their binary companions and environments in order to understand their progenitor systems.

At the position of SN~2006jc, a late-time source is significantly detected in the {\it WFC3/F300X} and {\it WFC3/F475X} bands. This source has a very stable brightness over 6.8 (or even 9.7) years since its last detection in 2010 observations. Detailed analysis rules out the possibility of a light echo or a new ejecta-CSM interaction at late times. Thus, we reinforce the conclusion by \citetalias{Maund2016} that it is most likely to be a binary companion of SN~2006jc's progenitor.

By fitting the stellar SED, we find that SN~2006jc's companion, with log($T_{\rm eff}$/K)~=~4.09$^{+0.05}_{-0.04}$ and log($L/L_\odot$)~=~4.52$^{+0.13}_{-0.13}$, is in the Hertzsprung gap and far from the MS. Further analysis suggests that it has experienced significant binary interaction with SN~2006jc's progenitor. The companion star has an initial mass of $M_2$~$\leq$~12.3$^{+2.3}_{-1.5}$~$M_\odot$, and the initial secondary-to-primary mass ratio is very close to unity ($q$~$\sim$~1). Thus, SN~2006jc may have had a not-so-massive progenitor, whose envelope was stripped by its binary companion.

We also discussed the possibility of obscuration by newly formed dust or of interaction with the SN ejecta. These scenarios may complicate the analysis of SN~2006jc's binary companion, and future observations will help to test these scenarios.

For SN~2015G, however, a companion star is not detected. This search was complicated by significant extinction towards the SN. We try to place an upper limit for its current mass with the detection limits and the BPASS models. If the companion is still close to the ZAMS, its upper mass limit can reach 60$^{+35}_{-25}$~$M_\odot$. At the coolest end [log($T_{\rm eff}$/K)~$\sim$~4.3], however, its current mass cannot exceed 27$^{+6}_{-5}$~$M_\odot$. It is also possible that SN~2015G may have a binary companion similar to that of SN~2006jc (which cannot be ruled out by the observations) or a compact-object companion.

We further analyse the environments of SNe~2006jc and 2015G. In both cases, the surrounding stellar populations are relatively old and argue against any massive WR stars as their progenitors. This also supports the conclusion that SNe~2006jc and 2015G have lower-mass progenitors arising from interacting binary systems.

Early work on SNe~Ibn has generally assumed that their progenitors are WR stars, which are initially massive and lose their envelopes via single-star evolution. This work suggests that the Type~Ibn SNe~2006jc and 2015G may actually come from lower-mass, interacting binary systems. It is quite surprising that even the class prototype SN~2006jc was produced via the binary progenitor channel. More investigation is needed to understand whether interacting binaries are a common progenitor channel for SNe~Ibn.

For SNe~2006jc and 2015G, we suggest that their pre-SN eruptions and CSM should have been produced by the progenitor stars themselves. Observations are not consistent with the recovered companion being an LBV. Thus, we reach a conclusion that pre-SN giant eruptions, which are commonly observed in massive ($M_{\rm ZAMS}~$~$>$~25~$M_\odot$) LBVs, can also occur in much lower-mass, stripped-envelope stars from interacting binaries. The previous speculation that WR stars may undergo such eruptions needs to be re-investigated, since it becomes increasingly unclear whether they could be SN~Ibn progenitors.

Thus, SNe~Ibn provide a unique opportunity to study the latest evolutionary stages of their stripped-envelope progenitors, when they may become wildly unstable and experience eruptive mass loss - something that has not been included in standard stellar evolutionary models. Combined with LBVs and SNe~IIn, SNe~Ibn allow us to investigate pre-SN eruptions in stars over a wider mass spectrum and with the presence/absence of hydrogen envelopes. We have discussed some possible mechanisms for pre-SN eruptions. More efforts are still needed in order to fully understand this intriguing phenomenon.

\section*{Acknowledgements}

We thank the anonymous referee for providing constructive comments to improve this paper. N.-C.S. thank Dr. Andrew Dolphin, Yi Yang, Chengyuan Li and Weijia Sun for their helpful suggestion in photometry and Emir Karamehmetoglu for discussion on light curve modelling. N.-C.S. acknowledge the funding support from the Science and Technology Facilities Council. The research of J.R.M. is funded through a Royal Society University Research Fellowship. R.H. was supported by the JSPS Overseas Research Fellowship No.29-514. This work is based on observations conducted with the {\it HST}, and the data was retrieved from the data archive at the Space Telescope Science Institute.

\appendix

\section{DOLPHOT Parameters}
\label{dp.sec}

\begin{figure}
\centering
\includegraphics[scale=0.55, angle=0]{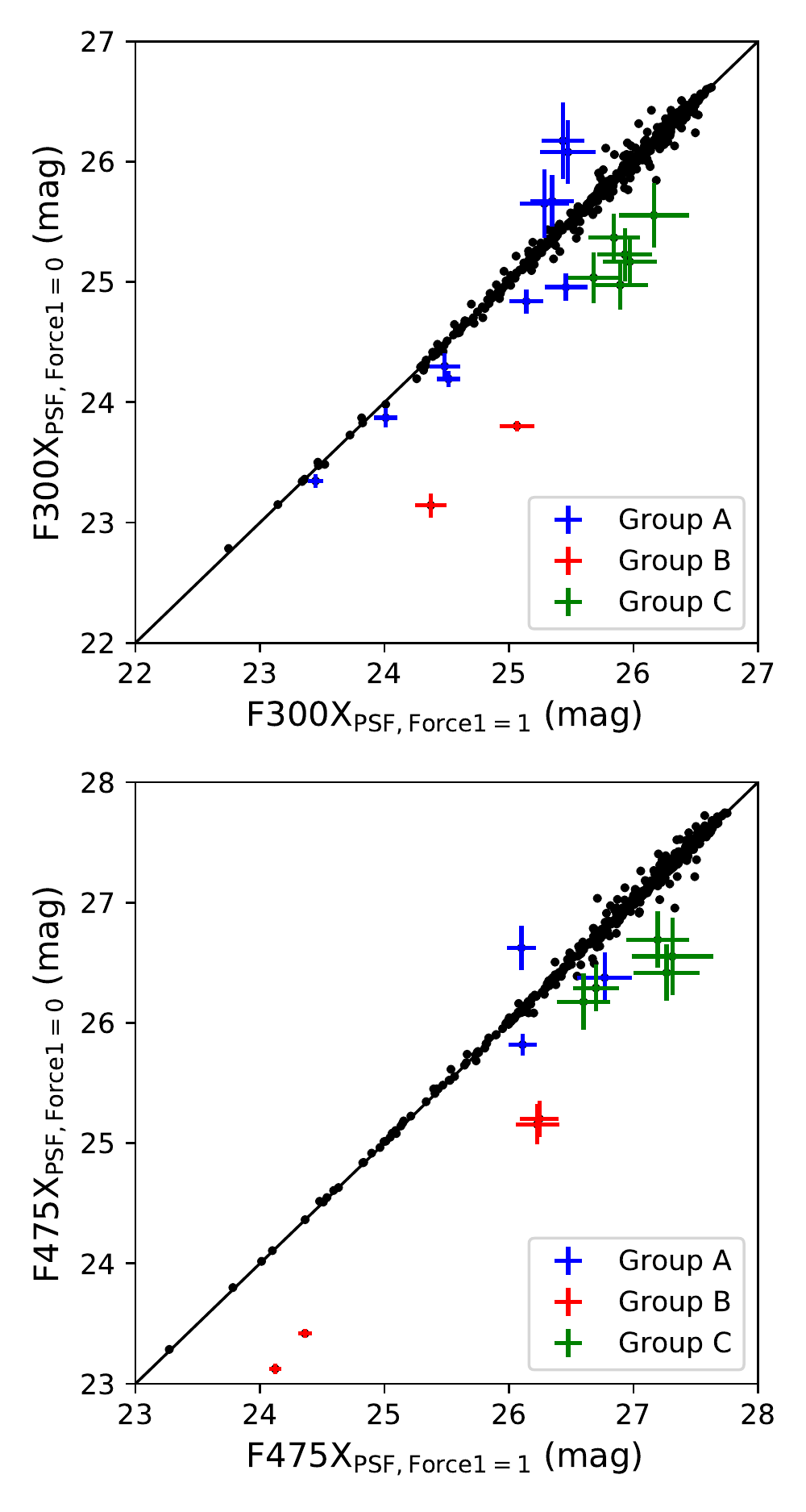}
\caption{Comparison of PSF photometry with \texttt{Force1~=~1} and \texttt{Force1~=~0} for the {\it WFC3} observations of SN~2006jc; black data points without error bars correspond to sources whose magnitude differences are within the photometric uncertainties; data points with error bars are those whose magnitude difference are larger than the photometric uncertainties, and they are further divided into three groups (A, blue; B, red; C, green; see text for details).}
\label{force1.fig}
\end{figure}

\begin{figure}
\centering
\includegraphics[scale=0.55, angle=0]{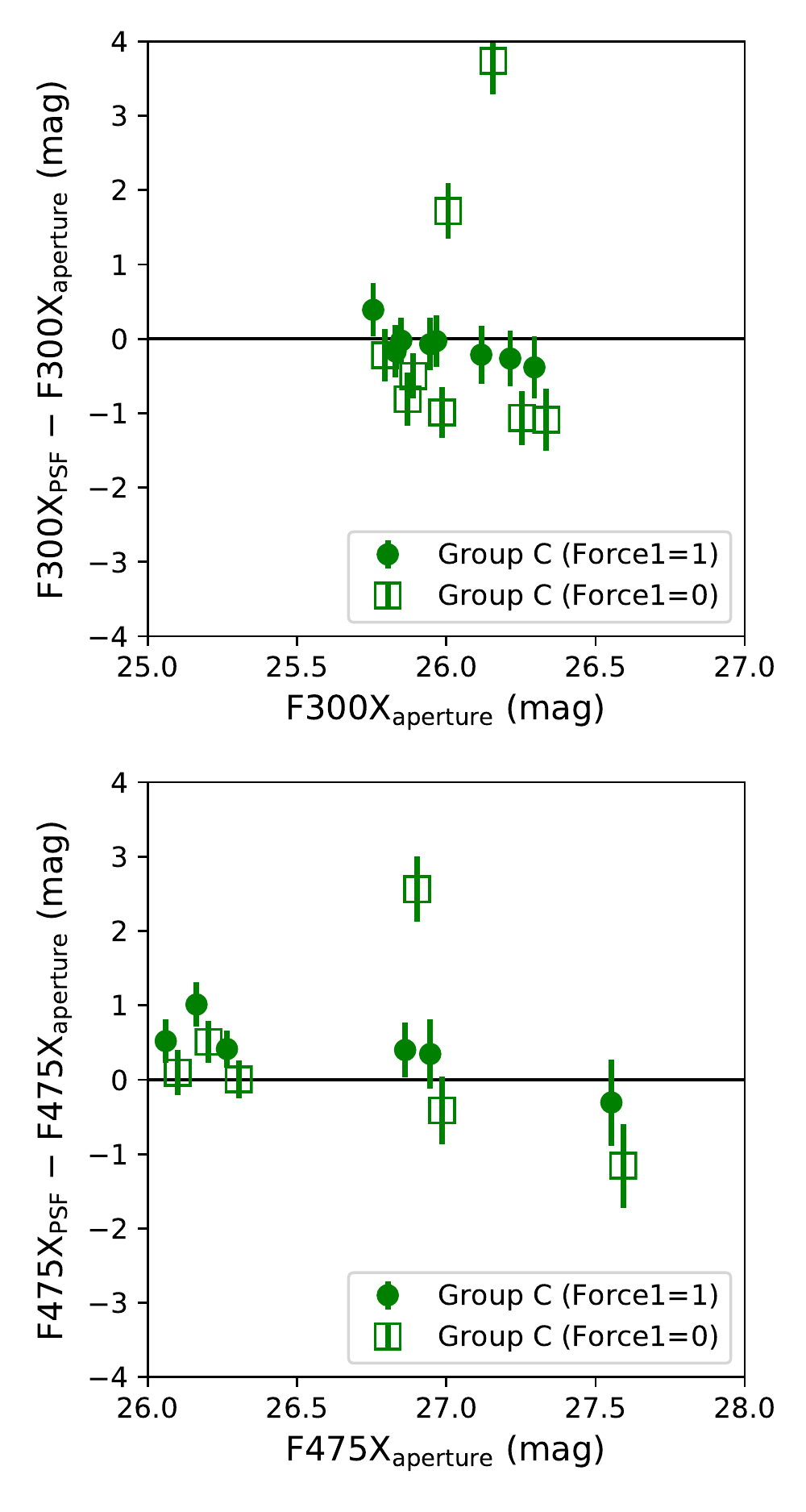}
\caption{Comparison of PSF photometry (\texttt{Force1~=~1}, filled circles; \texttt{Force1~=~0}, open squares) and aperture photometry for sources in Group~C; the vertical error bars reflect combined photometric uncertainties from PSF and aperture photometry; for clarity, the horizontal error bars have been omitted and small horizontal shifts ($\pm$~0.03~mag) have been applied to the data points.}
\label{force2.fig}
\end{figure}

As mentioned in Section~\ref{obs.sec}, we use DOLPHOT to conduct photometry of the observed {\it HST} images. DOLPHOT measures the magnitudes by fitting point spread functions (PSFs) to the detected sources. This process is controlled by a set of parameters, and changing the parameters may lead to different results. As a result, different groups may report inconsistent magnitudes even for the same source \citep[e.g. see Section~2 of][]{Eldridge2015}.

In this work, since both SNe occur in relatively sparse regions, we use \texttt{FitSky~=~3} (which controls the algorithm of sky fitting) and \texttt{Img\_RAper~=~8} (which sets the size of the aperture within which photometry will be performed), which are recommended by the user manual. We have tested the performance by changing to \texttt{FitSky~=~2} and \texttt{Img\_RAper~=~3}, which are optimal for crowded regions. The magnitudes are only slightly different and are consistent within the photometric uncertainties.

We find that setting \texttt{Force1~=~1} or \texttt{0} (force all objects to be fitted as stars or not) will cause a big difference in the photometry of some sources. For example, Fig.~\ref{force1.fig} shows a comparison of magnitudes derived with \texttt{Force1~=~1} and \texttt{Force1~=~0} for the {\it WFC3} observations of SN~2006jc (Program~14762). The magnitudes for most sources are quite consistent regardless of \texttt{Force1}. For a number of sources, however, the magnitude differences are larger than photometric uncertainties. We further find that these sources can be divided into three groups: (A) with object type parameters of TYPE~=~1 (good stars) or 2 (stars too faint for PSF determination), (B) TYPE~=~4 (too sharp) and relatively bright, and (C) TYPE~=~4 and relatively faint. Sources in Group~A are generally found in crowded regions; for these sources the results with \texttt{Force1~=~1} are more reliable since they are less affected by source crowding. Sources in Group~B are cosmic rays that have not been removed by ASTRODRIZZLE. Sources in Group~C are reported by DOLPHOT to be too sharp to be good stars. For these faint sources, however, the shape measurement may not be reliable: the extended wings of their PSFs may have been confused by the background, leaving only the PSF cores detectable.

In order to determine whether \texttt{Force1~=~1} or \texttt{0} should be used for Group~C, we further performed aperture photometry for these sources. Fig.~\ref{force2.fig} shows a comparison of PSF and aperture photometry. For {\it WFC3/F300X}, \texttt{Force1~=~1} gives results more consistent with aperture photometry. For {\it WFC3/F475X}, although results with \texttt{Force1~=~0} seem to have smaller deviations from aperture photometry at 26.0--26.5~mag, the agreement becomes much worse at fainter magnitudes. For SN~2006jc's binary companion, for instance, with \texttt{Force1~=~1} we obtain their magnitudes as $m_{F300X}$~=~25.93~$\pm$~0.22~mag and $m_{F475X}$~=~27.27~$\pm$~0.27~mag, but with \texttt{Force1~=~0} the magnitudes become $m_{F300X}$~=~25.23~$\pm$~0.22~mag, and $m_{F475X}$~=~26.42~$\pm$~0.23~mag. Aperture photometry gives $m_{F300X}$~=~26.22~$\pm$~0.27~mag and $m_{F475X}$~=~27.55~$\pm$~0.41~mag, which are consistent with the former set of magnitudes. Thus, we recommend using \texttt{Force1~=~1} in the PSF photometry.

Images from the {\it ACS} observations (Program~11675) have not been corrected for charge transfer inefficiency (i.e. they have filenames in the format of \texttt{*\_flt.fits}), and we apply an empirical correction to the measured magnitudes by setting \texttt{ACSuseCTE~=~1}. For the {\it WFC3} observations (Programs~14149 and 14762), the images have already been corrected for charge transfer inefficiency (i.e. with filenames in the format of \texttt{*\_flc.fits}). Thus, no additional correction is needed and we turned off that option by setting \texttt{WFC3useCTE~=~0}.

For the observations from Program~14762, we turned off aperture correction (\texttt{ApCor~=~0}) in DOLPHOT since there are not enough number of stars for this purpose. This does not cause any significant uncertainties since the the aperture correction is only a few hundredths of a magnitude (\citealt{Dolphin2000}; see also Section~4.3.2 of \citealt{Dalcanton2012}). For the other observations (Programs~11675 and 14149), aperture correction is performed by setting \texttt{ApCor~=~1}.

We use World Coordinate System (WCS) header information for alignment (\texttt{UseWCS~=~1}). All other parameters are the same as recommended by the DOLPHOT user manual.

\section{A Systematic Error}
\label{syserr.sec}

\begin{figure}
\centering
\includegraphics[scale=0.8, angle=0]{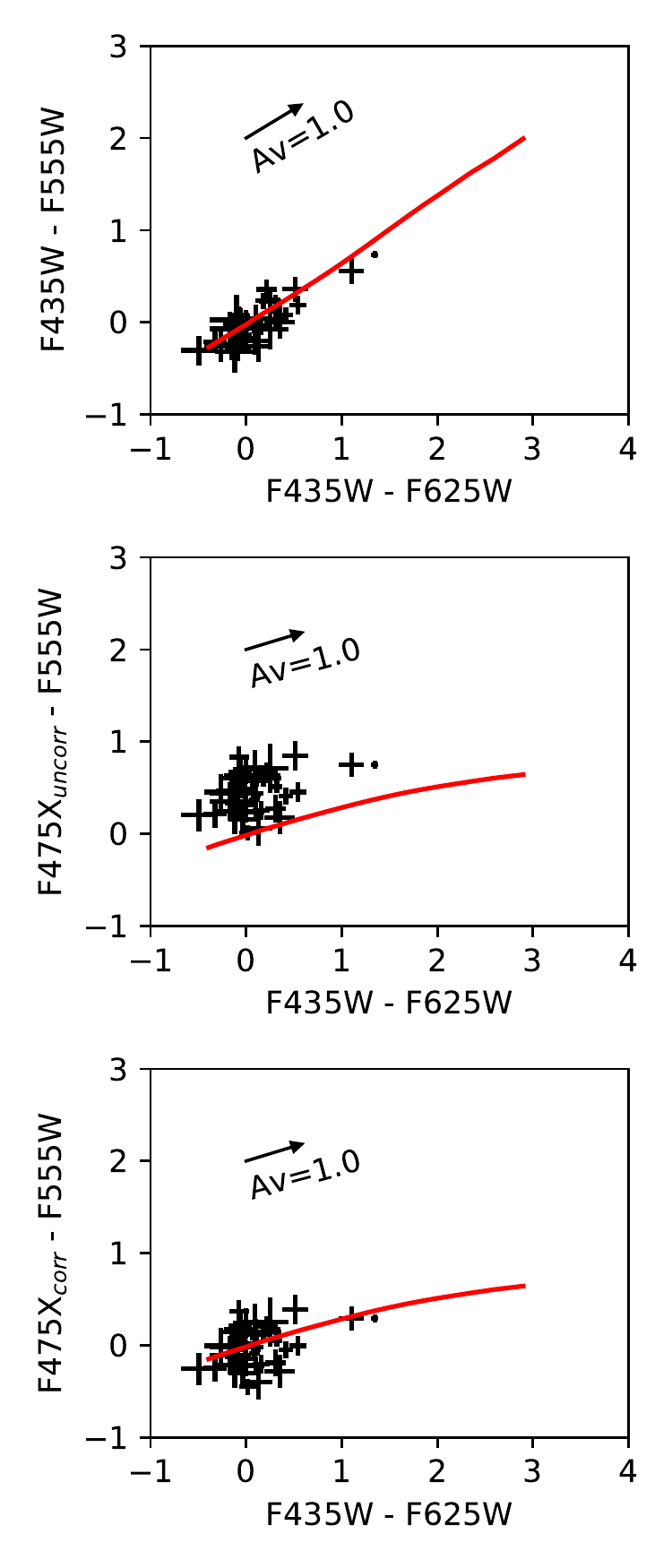}
\caption{Colour-colour diagrams of stars detected in the images of SN~2006jc. The errorbars reflect photometric uncertainties. The red lines are synthetic stellar locus from ATLAS9 \citep{Castelli2004} model spectra for supergiants, shifted by an interstellar reddening of $E(B-V)$~=~0.05~mag. In the middle panel, the {\it WFC3/F475X} magnitudes are from the raw outputs of DOLPHOT, while in the bottom panel  the {\it WFC3/F475X} magnitudes have been corrected by applying a shift of $-$0.46~mag. The arrow in each panel represents the reddening vector corresponding to $A_V$~=~1.0~mag.}
\label{ccd.fig}
\end{figure}

\begin{figure}
\centering
\includegraphics[scale=0.3, angle=0]{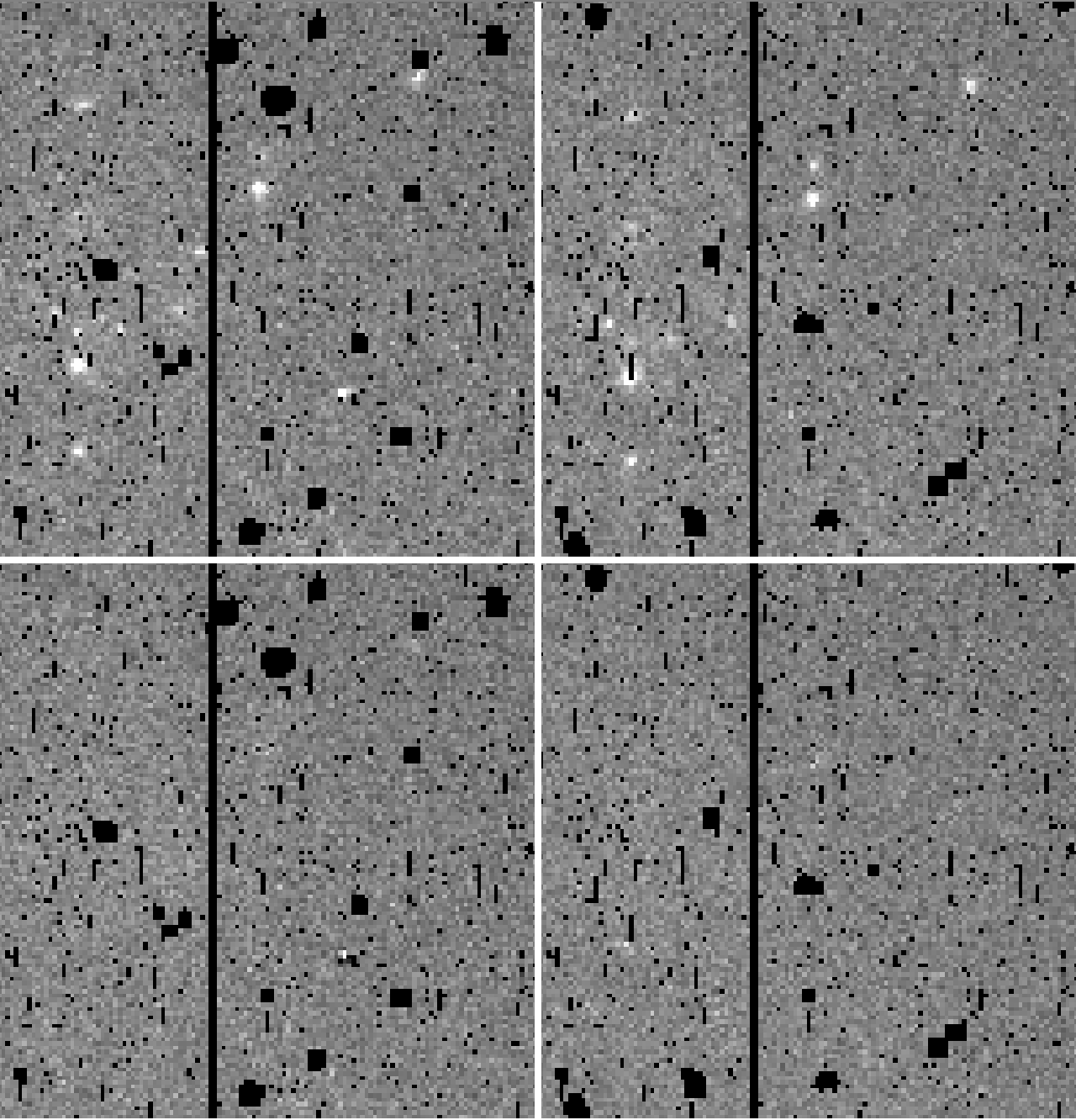}
\caption{Single-exposure {\it WFC3/F475X} images of SN~2006jc before (top row) and after (top row) PSF subtraction by DOLPHOT. The left/right column corresponds to the first/second exposure in this band. The black pixels in the images are bad pixels that have been masked in photometry. Only part of the full frame is shown for clarity.}
\label{res.fig}
\end{figure}

\begin{figure}
\centering
\includegraphics[scale=0.65, angle=0]{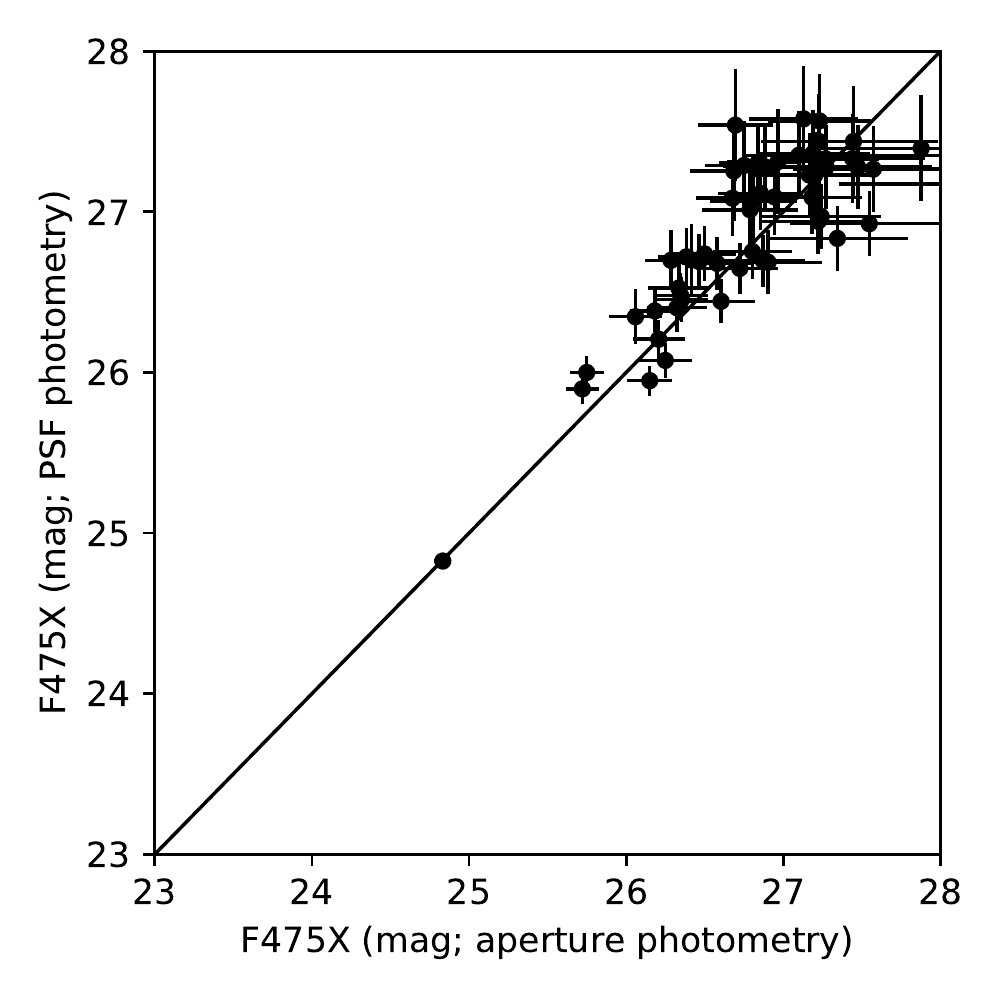}
\caption{Comparison of {\it WFC3/F475X} magnitudes between aperture photometry and PSF photometry for SN~2006jc.}
\label{apr.fig}
\end{figure}

\begin{figure}
\centering
\includegraphics[scale=0.8, angle=0]{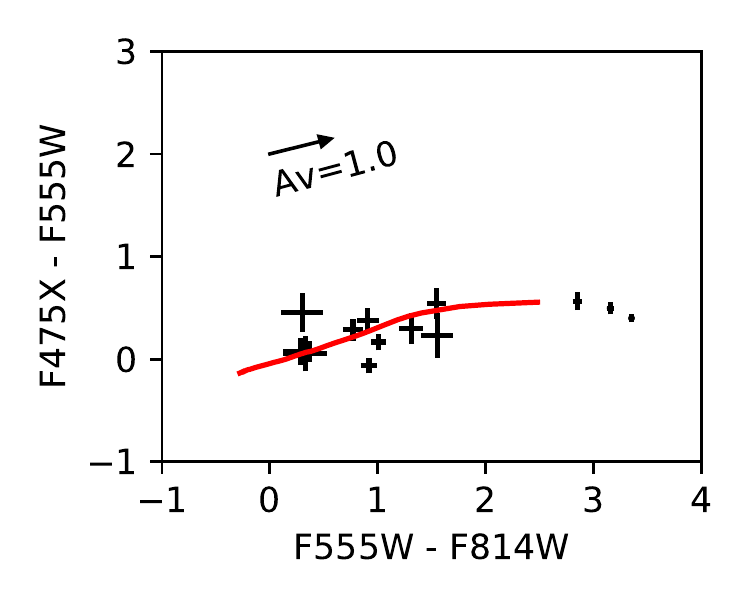}
\caption{Colour-colour diagrams of stars detected in the images of SN~2015G. The errorbars reflect photometric uncertainties. The red lines are synthetic stellar loci from ATLAS9 \citep{Castelli2004} model spectra for supergiants, shifted by an interstellar reddening of $E(B-V)$~=~0.384~mag. The {\it WFC3/F475X} magnitudes are from the raw outputs of DOLPHOT. The arrow in each panel represents the reddening vector corresponding to $A_V$~=~1.0~mag. The three sources with very red colours of {\it F555W - F814W}~$>$~2 are very bright ($m_{F814W}$~$\sim$~20--22~mag) and may be old globular clusters.}
\label{15g.fig}
\end{figure}

It is very important to check whether there is any potential systematic error in the photometric results, especially if one tries to compare the magnitudes at different epochs. When we did this we discovered a systematic error in the {\it WFC3/F475X} magnitudes in the observations of SN~2006jc (Program~14762). This is illustrated by the colour-colour diagrams (Fig.~\ref{ccd.fig}, in which stars detected in different bands are cross-matched with a conservative matching radius of 0.3~pixel). In the top panel, colours with the {\it ACS/F435W}, {\it ACS/F555W}, and {\it ACS/F625W} filters match well with the expectation of the theoretical stellar locus. Thus, we believe that photometry in these bands should be reliable. In the middle panel, however, the {\it F475X}~$-$~{\it F555W} colour is systematically redder than the stellar locus.

This mismatch cannot be explained by interstellar reddening since the reddening vector is almost parallel to the stellar locus. This mismatch is also unlikely to be caused by incorrect stellar locus. Note that {\it WFC3/F475X} and {\it WFC3/F555W} have similar wavelength coverages, with only a difference of 500~\AA\ in their short-wavelength cut-off. Thus, hot sources should have {\it F475X}~$-$~{\it F555W} colours close to zero in the Vega magnitude system (as both bands cover the Rayleigh-Jeans tail of their SEDs). Indeed, the {\it F475X}~$-$~{\it F555W} colour is very close to zero at the blue end of the stellar locus. Furthermore, the synthetic magnitudes of the stellar locus has been independently calculated by two authors (one with the Python package PySynphot\footnote{\url{https://pysynphot.readthedocs.io}} and the other with his own code), which are consistent with each other. Our calculated stellar locus is also consistent with that from the PARSEC \citep[v1.2S;][]{Bressan2012} models.

We have further checked the correctness of the {\it WFC3/F475X} photometry. Figure~\ref{res.fig} shows the {\it WFC3/F475X} images of SN~2006jc before and after PSF subtraction by DOLPHOT. The residual images look very clean after PSF subtraction; thus, the PSF modelling has been performed very well by DOLPHOT for this dataset. We also note that the built-in photometric zeropoints are consistent with those published by the Space Telescope Science Institute\footnote{\url{http://www.stsci.edu/hst/instrumentation/wfc3/data-analysis/photometric-calibration/uvis-photometric-calibration}}. In addition, we have selected a number of isolated stars to perform aperture photometry. Figure~\ref{apr.fig} shows a comparison between the PSF and aperture photometry, where no systematic discrepancy is found. Thus, the {\it WFC3/F475X} photometry should be reliable. We have also requested a colleague to do photometry independently, but the deviation from the stellar locus is still apparent in his results.

We have also checked the {\it WFC3/F475X} magnitudes for SN~2015G, the colour-colour plot of which is displayed in Fig.~\ref{15g.fig}. For this dataset, however, the {\it F475X}~$-$~{\it F555W} colours are in agreement with the theoretical stellar locus without any systematic deviations.

To estimate the systematic error in {\it WFC3/F475X} magnitudes for SN~2006jc, we first fit a 3-order polynomial function to the theoretical stellar locus ({\it F475X}~$-$~{\it F555W} v.s. {\it F435W}~$-$~{\it F625W}). Next, we predict the ``theoretical" {\it F475X}~$-$~{\it F555W} colours of the detected stars by applying this function to their {\it F435W}~$-$~{\it F625W} colours. Then we calculate the differences between the ``theoretical" and the measured {\it F475X}~$-$~{\it F555W} colours for all detected stars. Finally, the systematic error is estimated by their inverse-variance average with weights coming from photometric errors.

We find this systematic error to be 0.46~mag and we choose to subtract this value from the {\it WFC3/F475X} magnitudes from all sources in the observation of SN~2006jc. The bottom panel of Fig.~\ref{ccd.fig} shows a colour-colour diagram with the corrected {\it WFC3/F475X} magnitudes, in which the stars agree well with the theoretical locus. For SN~2006jc's binary companion, the magnitude reported by DOLPHOT is $m_{F475X}$~=~27.27~$\pm$~0.27~mag, and after applying this shift it becomes $m_{F475X}$~=~26.81~$\pm$~0.27~mag.

\bsp	
\label{lastpage}
\end{document}